\documentclass[oldversion]
{aa}

\usepackage{amsmath}
\usepackage{natbib}
\usepackage{epsfig}
\usepackage{txfonts}

\newcommand{\id}{{\,\rm d}}

\newcommand{\beq}{\begin{equation}}   %

\newcommand{\eeq}{\end{equation}}   %

\newcommand{\beqa}{\begin{eqnarray}}   %

\newcommand{\eeqa}{\end{eqnarray}}   %

\newcommand{\beal}{\begin{align}}

\newcommand{\enal}{\end{align}}

\newcommand{\bspl}{\begin{split}}

\newcommand{\espl}{\end{split}}

\newcommand{\bsub}{\begin{subequations}}

\newcommand{\esub}{\end{subequations}}

\newcommand{\bmulti}{\begin{multline}}   %

\newcommand{\beqm}{\begin{mathletters}}   %

\newcommand{\eeqm}{\end{mathletters}}   %

\newcommand{\Abst}[1]{\,#1}

\newcommand{\kB}{k_{\rm B}}

\newcommand{\me}{m_{\rm e}}

\newcommand{\Ne}{N_{\rm e}}

\newcommand{\Np}{N_{\rm p}}

\newcommand{\Te}{T_{\rm e}}

\newcommand{\Tg}{T_{\gamma}}

\newcommand{\sigT}{\sigma_{\rm T}}

\newcommand{\pd}{\partial}

\newcommand{\pAb}[2]{\frac{\displaystyle\pd #1}{\displaystyle\pd #2}}

\newcommand{\pot}[2]{#1 \times 10^{#2}}

\newcommand{\change}[1]{{#1}}

\newcommand{\zh}{{z_{\rm h}}}
\newcommand{\Nc}{{N_{\rm c}}}
\newcommand{\NHeIII}{{N_{\ion{He}{iii}}}}
\newcommand{\nuic}{{\nu_{i\rm c}}}
\newcommand{\xic}{{x_{i\rm c}}}
\newcommand{\xtwopc}{{x_{\rm 2p c}}}
\newcommand{\xonesc}{{x_{\rm 1s c}}}

\newcommand{\ye}{{y_{\rm e}}}

\newcommand{\zem}{{z_{\rm em}}}
\newcommand{\sigav}[2]{{\left<\, #1 \,\right>_{#2}}}
\newcommand{\Nloop}{{\mathcal{N}_{\rm loop}}}
\newcommand{\NloopH}{{\mathcal{N}^{\ion{H}{i}}_{\rm loop}}}
\newcommand{\NloopHe}{{\mathcal{N}^{\ion{He}{ii}}_{\rm loop}}}

\begin{document}

%

\titlerunning{Pre-recombinational energy release and narrow features in the CMB spectrum}
\title{Pre-recombinational energy release and \\ narrow features in the CMB spectrum}

\author{J. Chluba\inst{1} \and R.A. Sunyaev\inst{1,2}}
\authorrunning{Chluba \and Sunyaev}

\institute{Max-Planck-Institut f\"ur Astrophysik, Karl-Schwarzschild-Str. 1,
85741 Garching bei M\"unchen, Germany 
\and 
Space Research Institute, Russian Academy of Sciences, Profsoyuznaya 84/32,
117997 Moscow, Russia
}

\offprints{J. Chluba, 
\\ \email{jchluba@mpa-garching.mpg.de}
}

\date{Received / Accepted}

\abstract{Energy release in the early Universe ($z\lesssim \pot{2}{6}$) should
  lead to some broad spectral distortion of the cosmic microwave background
  (CMB) radiation field, which can be characterized as $y$-type distortion
  when the injection process started at redshifts $z\lesssim \pot{5}{4}$.
  Here we demonstrate that if energy was released before the beginning of
  cosmological hydrogen recombination ($z\sim 1400$), closed loops of
  bound-bound and free-bound transitions in \ion{H}{i} and \ion{He}{ii} lead
  to the appearance of (i) characteristic multiple narrow spectral features at
  dm and cm wavelengths, and (ii) a prominent sub-millimeter feature
  consisting of absorption and emission parts in the far Wien tail of CMB
  spectrum.
	\change{The additional spectral features are generated in the
	pre-recombinational epoch of \ion{H}{i} ($z\gtrsim 1800$) and
	\ion{He}{ii} ($z\gtrsim 7000$), and therefore differ from those arising
	due to normal cosmological recombination in the undisturbed CMB blackbody
	radiation field.}
  We present the results of numerical computations including 25 atomic shells
  for both \ion{H}{i} and \ion{He}{ii}, and discuss the contributions of
  several individual transitions in detail.
As examples, we consider the case of instantaneous energy release (e.g. due to
phase transitions) and exponential energy release \change{because of}
long-lived decaying particles.
\change{Our computations show that due to possible pre-recombinational atomic
transitions the variability of the CMB spectral distortion increases when
comparing with the distortions arising in the normal recombination epoch.}
\change{The amplitude of the spectral features, both at low and high
frequencies, directly depends on the value of the $y$-parameter, which
describes the intrinsic CMB spectral distortion resulting from the energy
release. Also the time-dependence of the injection process play an important
role, for example leading to non-trivial shifts of the quasi-periodic pattern
at low frequencies along the frequency axis.}
The existence of these narrow spectral features would open an unique way to
separate $y$-distortions due to pre-recombinational ($1400\lesssim z \lesssim
\pot{5}{4}$) energy release from those arising in the post-recombinational era
at redshifts $z\lesssim 800$.
}       

\maketitle

\section{Introduction}
\label{sec:Intro}
The measurements with the {\sc Cobe/Firas} instrument have proven that the
spectrum of the cosmic microwave background (CMB) is very close to a perfect
blackbody \citep{Fixsen1996} with thermodynamic temperature $T_0=2.725\pm
0.001\,$K \citep{Mather1999, Fixsen2002}.
However, from the theoretical point of view deviations of the CMB spectrum
from the one of a pure blackbody are not only possible but even inevitable if,
for example, energy was released in the {\it early Universe} (e.g. due to
viscous damping of acoustic waves, or annihilating or decaying
particles). For very early energy release ($\pot{5}{4}\lesssim z \lesssim
\pot{2}{6}$) the resulting spectral distortion can be characterized as a
Bose-Einstein $\mu$-type distortion \citep{Sunyaev1970b, Illarionov1975a,
  Illarionov1975b}, while for energy release at low redshifts ($z\lesssim
\pot{5}{4}$) the distortion is close to a $y$-type distortion
\citep{Zeldovich1969}.
The current best observational limits on these types of distortions
are $|y|\leq \pot{1.5}{-5}$ and $|\mu|\leq \pot{9.0}{-5}$ \citep{Fixsen1996}.
Due to the rapid technological development, improvements of these limits by a
factor of $\sim 50$ in principle could have been possible already several
years ago \citep{Fixsen2002}, and recently some efforts are made to determine
the absolute value of the CMB brightness temperature at low frequencies
using {\sc Arcade} \citep{Kogut2004, Kogut2006}.
%

Also in the {\it post-recombinational epoch} ($z\lesssim 800$), $y$-type
spectral distortions due to different physical mechanisms should be produced.
Performing measurements of the average CMB spectrum (e.g. with wide-angle
horns or like it was done with {\sc Cobe/Firas}) all clusters of galaxies,
hosting hot intergalactic gas, due to the thermal SZ-effect
\citep{Sunyaev1972} are contributing to the integral value of the observed
$y$-parameter.
Similarly supernova remnants at high redshifts \citep{Oh2003}, or shock waves
arising due to large-scale structure formation \citep{Sunyaev1972b, Cen1999,
Miniati2000} should contribute to the overall \change{$y$-parameter.
For its value} today we only have the upper limit by {\sc Cobe/Firas},
and lower limits derived by estimating the total contribution of all clusters
in the Universe \citep{Markevitch1991, Silva2000, Roncarelli2007}.
These lower limits are exceeding $y\sim 10^{-6}$, and it is still possible
  that the contributions to the total value of $y$ due to early energy release
  are comparable or exceeding those coming from the low redshift Universe.

Several detailed analytical and numerical studies for various 
energy injection histories and mechanisms can be found in the literature
\citep[e.g.][]{Zeldovich1969, Sunyaev1970b, Sunyaev1970COA,
Sunyaev1970Antimatter, Illarionov1975a, Illarionov1975b, Zeldovich1972,
Chan1975, Danese1982, Daly1991, Burigana1991, Burigana1991b, Burigana1995,
Hu1993, Hu1993a, Hu1994, Salvaterra2002, Burigana2003, Chluba2004}.
Two very important conclusions can be drawn from these all studies: (i) the
arising spectral distortions are always {\it very broad} and practically
{\it featureless}, and (ii) due to the absence of narrow spectral features,
distinguishing different injection histories is extremely difficult.
This implies that if one would find a $y$-type spectral distortion in the
average CMB spectrum, then it is practically impossible to say if the energy
injection occurred just before, during or after the epoch of cosmological
recombination.
In this paper we show that the {\it pre-recombinational emission} within the
bound-bound and free-bound transition of atomic hydrogen and helium should
leave multiple {\it narrow features} ($\Delta\nu/\nu\sim 10-30\%$) in the CMB
spectrum, that might become observable at cm, dm and sub-mm wavelength (see
Sect.~\ref{sec:results}). This could in principle open a way to directly
distinguish pre- and post-recombinational $y$-distortions and even to shed
light on the time-dependence of the energy injection process.

At redshifts well before the \change{epoch of $\ion{He}{iii}\rightarrow
\ion{He}{ii}$-recombination} ($z~\gtrsim~8000$) the total number of CMB
photons is not affected by atomic transitions if the intrinsic CMB spectrum is
given by a {\it pure blackbody}.
This is because the atomic emission and absorption processes in full
thermodynamic equilibrium balance each other.
However, a \change{lower} redshifts ($z~\lesssim~8000$), due to the expansion
of the Universe, the medium became sufficiently cold to allow the formation of
neutral atoms.
The transition to the neutral state is associated with the release of several
additional photons per baryon (e.g. $\sim 5$ photons per hydrogen atom
\citep{Chluba2006b}), even within a pure blackbody ambient CMB radiation
field.
\change{Refining early estimates \citep{Zeldovich68, Peebles68, Dubrovich1975,
    DubroVlad95, Dubrovich1997},} the spectral distortions arising during
    hydrogen recombination ($800\lesssim z \lesssim 1800$),
    $\ion{He}{ii}\rightarrow \ion{He}{i}$-recombination ($1600\lesssim z
    \lesssim 3000$), and $\ion{He}{iii}\rightarrow \ion{He}{ii}$-recombination
    ($4500\lesssim z \lesssim 7000$) \change{within a pure blackbody ambient
    radiation field} have been recently computed in detail \citep{Jose2006,
    Chluba2006b, Chluba2007, Jose2007}. It was also emphasized that measuring
    these distortions in principle may open another independent way to
    determine the temperature of the CMB monopole, the specific entropy of the
    Universe, and the primordial helium abundance, well before the first
    appearance of stars \citep[e.g.][]{RS2007, Chluba2008, RS2008}.

If on the other hand the intrinsic CMB spectrum deviates from a pure
blackbody, full equilibrium is perturbed, and the small imbalance between
emission and absorption in atomic transitions can lead to a net change of the
number of photons, even prior to the epoch of recombination, \change{in
particular} owing to {\it loops} starting and ending in the continuum
\citep{Liubarskii83}.
These loops are trying to diminish the maximal spectral distortions and are
producing several new photons per absorbed one.
In this paper we want to demonstrate how the cosmological recombination
spectrum is affected if one allows for an intrinsic $y$-type CMB spectral
distortion. We investigate the cases of single instantaneous energy injection
(e.g. due to phase transitions) and for long-lived decaying particles.
There is no principle difficulty in performing the calculations for more
general injection histories, also including $\mu$-type distortions, if
necessary. However, this still requires a slightly more detailed study, which
will be left for a future paper.
%

%
In Sect.~\ref{sec:energy_dist} we provide a short overview regarding the
thermalization of CMB spectral distortions after early energy release, and
provide \change{formulae which we then use in our computations to describe
$y$-type distortions.}
In Sect.~\ref{sec:atom_trans} we give explicit expressions for the net
bound-bound and free-bound rates in a distorted ambient radiation field. 
We then derive some estimates for the expected contributions to the
pre-recombinational signals coming from primordial helium in
Sect.~\ref{sec:helium_cont}.
Our main results are presented in Sect.~\ref{sec:results}, where we first
start by discussing a few simple cases (Sect.~\ref{sec:2shellresults} and
\ref{sec:3shellresults}) in order to gain some level of understanding.
We support our numerical computations by several analytic considerations in
Sect~\ref{sec:Lya2shellresults} and Appendix~\ref{app:analytic}.
In Sect.~\ref{sec:25shellresults} we then discuss the results for our 25 shell
computations of hydrogen and \ion{He}{ii}. First we consider the dependence of
the spectral distortions on the value of $y$ (Sect.~\ref{sec:ydep}), where
Fig.~\ref{fig:HI_HeII_y} and \ref{fig:HI.HeII.contributions.vh} play the main
role.
Then in Sects.~\ref{sec:zdep} and \ref{sec:yzdep} we investigate the
dependence of the spectral distortions on the injection redshift and history,
where we are particularly interested in the low frequency variability of the
signal (see Fig.~\ref{fig:DT.25.diff_z} and \ref{fig:DT.25.diff_z.decay}).
We conclude in Sect.~\ref{sec:conc}.

\section{CMB spectral distortions after energy release}
\label{sec:energy_dist}
After any energy release in the Universe, the thermodynamic
equilibrium between matter and radiation in general will be perturbed, and in
particular, the distribution of photons will deviate from the one of a pure
blackbody.
The combined action of Compton scattering, double Compton
emission\footnote{Due to the huge excess in the number of photons over baryons
($N_\gamma/N_{\rm b}\sim \pot{1.6}{9}$), the double Compton process is the
dominant source of new photons at redshifts $z_{\rm dc}\gtrsim
\pot{3}{5}-\pot{4}{5}$, while at $z\lesssim z_{\rm dc}$ Bremsstrahlung is more
important.} \citep{Lightman1981, Thorne1981, Chluba2007DC}, and Bremsstrahlung
will attempt to restore full equilibrium, but, depending on the injection
redshift, might not fully succeed.
Using the approximate formulae given in \citet{Burigana1991} and
\citet{Hu1993}, for the parameters within the concordance cosmological model
\citep{Spergel2003, WMAP_params}, one can distinguish between the following cases for the
residual CMB spectral distortions arising from a single energy injection,
$\delta\rho_\gamma/\rho_\gamma\ll 1$, at heating redshift $z_{\rm h}$:

\begin{enumerate}
\item[(I)] $\zh< z_{y}\sim \pot{6.3}{3}$: Compton scattering is {\it not able}
to establish full kinetic equilibrium of the photon distribution with the
electrons. Photon producing processes (mainly Bremsstrahlung) can only restore
a Planckian spectrum at very low frequencies.  Heating results in a Compton
$y$-distortion \citep{Zeldovich1969} at high frequencies, like in the case of
the thermal SZ effect, with $y$-parameter
$y\sim\frac{1}{4}\,\delta\rho_\gamma/\rho_\gamma$.
  
\item[(II)] $z_{y}<\zh<z_{\mu}\sim \pot{2.9}{5}$: Compton scattering can
establish {\it partial} kinetic equilibrium of the photon distribution with
the electrons. Photons that are produced at low frequencies (mainly due to
Bremsstrahlung) diminish the spectral distortion close to their initial
frequency, but cannot upscatter strongly. The deviations from a blackbody
represent a {\it mixture} between a $y$-distortion and a $\mu$-distortion.
  
\item[(III)] $z_{\mu}<\zh< z_{\rm th}\sim \pot{2}{6}$: Compton scattering can
establish {\it full} kinetic equilibrium of the photon distribution with the
electrons after a very short time. Low frequency photons (mainly due to double
Compton emission) upscatter and slowly reduce the spectral distortion at high
frequencies
%
%
The deviations from a blackbody can be described as a Bose-Einstein
distribution with frequency-dependent chemical potential, which is constant at
high and vanishes at low frequencies.
  
\item[(IV)] $z_{\rm th}<\zh$: Both Compton scattering and photon production
processes are extremely efficient and restore practically any spectral
distortion arising after heating, eventually yielding a pure blackbody
spectrum with slightly higher temperature $\Tg$ than before \change{the energy
release}.
 
\end{enumerate}

For the case (I) and (III) it is possible to approximate the distorted radiation
spectrum analytically. There is no principle difficulty in numerically
computing the time-dependent solution for the radiation field after release of
energy \citep[e.g.][]{Hu1993} for more general cases, if necessary.
However, here we are particularly interested in demonstrating the main
difference in the additional radiation appearing due to atomic transitions in
hydrogen and helium {\it before} the actual epoch of recombination, and for
late energy release, in which a $y$-type spectral distortion is formed.
We will therefore only distinguish between case (I) and (III), and assume that
the transition between these two cases occurs at $z_{\mu,y}\approx z_{\mu}/4
\sqrt{2}\sim \pot{5.1}{4}$, i.e. the redshift at which the energy exchange
time scale equals the expansion time scale \citep{Sunyaev1980ARAA, Hu1993}.
The effects due to intrinsic $\mu$-type spectral distortions will be left for
a future work, so that below we will restrict ourselves to energy injection at
redshift $z\lesssim 50000$.

\subsection{Compton $y$-distortion}
\label{sec:ydist}
For energy release at low redshifts the Compton process is no longer able to
establish kinetic equilibrium.
If the temperature of the radiation is smaller than the temperature of the
electrons photons are upscattered. For photons which are initially distributed
according to a blackbody spectrum with temperature $\Tg$, the efficiency of
this process is determined by the Compton $y$-parameter,
\beal
\label{eq:y_par_Comp}
y=\int \frac{\kB(\Te-\Tg)}{\me c^2}\,\Ne\,\sigT \id l
\Abst{,}
\end{align}
where $\sigT$ is the Thomson cross section, $\id l=c\id t$, $\Ne$ the electron number density,
and $\Te$ the electron temperature.
For $y\ll 1$, the resulting intrinsic distortion in the photon occupation
number of the CMB is approximately given by \citep{Zeldovich1969}
\beal
\label{eq:y_dist}
\Delta n_{\gamma}
=y\,\frac{x\,e^{x}}{(e^{x}-1)^2}\left[x\,\frac{e^{x}+1}{e^{x}-1}-4\right]
\Abst{.}
\end{align}
Here $x=h\nu/\kB \Tg$ is the dimensionless frequency.

For computational reasons it is convenient to introduce the frequency
dependent chemical potential resulting from a $y$-distortion, which can be
obtained with
\beal
\label{eq:mu_n}
\mu(x)&=\ln\left(\frac{1+n_\gamma}{n_\gamma}\right)-x
\stackrel{\stackrel{y\ll 1}{\downarrow}}{\approx}
-y\,x\,\left[x\,\frac{e^{x}+1}{e^{x}-1}-4\right]
\Abst{.}
\end{align}
Here $n_{\rm pl}(x)=1/[e^x-1]$ is the Planckian occupation number, and
$n_\gamma=n_{\rm pl}+\Delta n_{\gamma}$. For $x\rightarrow 0$ and $y\ll 1$ one
finds $\mu(x)\approx 2x y$, and for $x\gg 1$ one has $\mu(x)\approx
-\ln[1+y\,x^2]$, or $\mu(x)\approx -y\,x^2$ for $1\ll x\ll \sqrt{1/y}$.
Comparing with a blackbody spectrum of temperature $\Tg$, for $y>0$ there is a
deficit of photons at low frequencies, while there is an excess at high
frequencies.
In particular, the spectral distortion changes sign at $x_y\sim 3.8$. 

\subsubsection{Compton $y$-distortion from decaying particles}
\label{sec:ydist_decay}
If all the energy is released at a single redshift, $z_i\lesssim z_{\mu,y}\sim
50000$, then after a very short time a $y$-type distortion is formed, where
the $y$-parameter is approximately given by $y\sim\frac{1}{4}\,\delta
\rho_\gamma/\rho_\gamma$.

However, when the energy release is due to decaying unstable particles, which
have sufficiently long life-times, $t_{\rm X}$, then the CMB spectral
distortion will built up as a function of redshift. In this case the
fractional energy injection rate is given by $\delta
\dot{\rho}_\gamma/\rho_\gamma\propto e^{-t(z)/t_{\rm X}}/(1+z)$, so that the
time-dependent $y$-parameter can be computed as
\beal
\label{eq:y_z_decay}
y(z)=
y_{0}\times\frac{\int_z^{\infty}  \id z'\,e^{-t(z')/t_{\rm X}}/H(z')(1+z')^2}
{\int_0^{\infty}  \id z'\,e^{-t(z')/t_{\rm X}}/H(z')(1+z')^2}
\Abst{,}
\end{align}
where $y_{0}=\frac{1}{4}\,\delta \rho_\gamma/\rho_\gamma$ is related to the
total energy release, and $H(z)$ is the Hubble expansion factor. Note that
$y(z)$ is a rather steep function of redshift, which strongly rises around the
redshift, $z_{\rm X}$, at which $t(z)\equiv t_{\rm X}$.

\section{Atomic transitions in a distorted ambient CMB radiation field}
\label{sec:atom_trans}

\subsection{Bound-bound transitions}
\label{sec:bb_trans}
Using the occupation number of photons, $n_\gamma=1/[e^{x+\mu}-1]$, with
frequency-dependent chemical potential $\mu(x)$, one can express the net rate
connecting two bound atomic states $i$ and $j$ in the convenient form
\beal
\label{eq:DRij_mu}
\Delta
R_{ij}=p_{ij}\,\frac{A_{ij}\,N_i\,e^{x_{ij}+\mu_{ij}}}{e^{x_{ij}+\mu_{ij}}-1}
\left[1-\frac{g_i}{g_j}\,\frac{N_j}{N_i}\,e^{-[x_{ij}+\mu_{ij}]}\right]
\Abst{,}
\end{align}
where $p_{ij}$ is the Sobolev-escape probability, $A_{ij}$ is the
Einstein-$A$-coefficient of the transition $i\rightarrow j$, $N_i$ and $g_i$
are the population and statistical weight of the upper and $N_j$ and $g_j$ of
the lower hydrogen level, respectively. Furthermore we have introduced the
dimensionless frequency $x_{ij}=h\nu_{ij}/k T_0 (1+z)$ of the transition,
where $T_0=2.725\,$K is the present CMB temperature \citep{Fixsen2002}, and
$\mu_{ij}=\mu(x_{ij})$.

\subsection{Free-bound transitions}
\label{sec:fb_trans}
For the free-bound transitions from the continuum to the bound atomic states $i$
one has
\beal
\label{eq:DRci_mu}
\Delta
R_{{\rm c}i}=\Ne\,\Nc\,\alpha_{i}-N_i\,\beta_i
\Abst{,}
\end{align}
where $\Nc$ in the case of hydrogen is the number density of free protons,
$\Np$, and the number density for \ion{He}{iii} nuclei, $\NHeIII$, in the case
of helium. The recombination coefficient, $\alpha_{i}$, and photoionization
coefficient, $\beta_i$, are given by the integrals
\bsub
\label{eq:alphabetai}
\beal
\alpha_{i}&=\frac{8\pi}{c^2}\,\tilde{f}_i(\Te)\,\int_{\nuic}^\infty
\frac{\nu^2\,\sigma_i(\nu)\,e^{x+\mu(x)+(\xic-x)/\rho}}{e^{x+\mu(x)}-1}\id\nu
\\
\beta_i&=\frac{8\pi}{c^2}\,\int_{\nuic}^\infty\frac{\nu^2\sigma_i(\nu)}{e^{x+\mu(x)}-1}\id\nu
\Abst{,}
\end{align}
\esub
Here $\xic=h\nuic/k\Tg$ is the dimensionless ionization frequency,
$\rho=\Te/\Tg$ is the ration of the photon and electron temperature,
$\sigma_i$ is the photoionization cross-section for the level $i$, and
%
$\tilde{f}_i(\Te)=\frac{g_i}{2}\,\left[\frac{h^2}{2\pi\me
k\Te}\right]^{3/2}\approx\frac{g_i}{2}\,\pot{4.14}{-16}\,T^{-3/2}_{\rm
e}\,{\rm cm^3}$.
In full thermodynamic equilibrium the photon distribution is given by a
blackbody with $\Tg=\Te$. As expected, in this case one finds from
Eq.~\eqref{eq:alphabetai} that $\alpha^{\rm
  eq}_i\equiv \tilde{f}_i(\Te)\,e^{h\nuic/k\Te}\,\beta^{\rm eq}_i$.

\section{Expected contributions from helium}
\label{sec:helium_cont}
The number of helium nuclei is only $\sim 8\%$ relative to the number of
hydrogen atoms in the Universe. Compared to the radiation coming from hydrogen
one therefore naively expects a small addition of photons due to atomic
transitions in helium.
However, at given frequency the photons due to \ion{He}{ii} have been released
at about $Z^2=4$ times higher redshifts than for hydrogen, so that both the
number density of particles and temperature of the medium was higher. In
addition the expansion of the Universe was faster.
As we will show below, these circumstances make the contributions from helium
comparable to those from hydrogen, where \ion{He}{ii} plays a much more
important role than \ion{He}{i}.

\subsection{Contributions due to \ion{He}{ii}}
The speed at which atomic loops can be passed through is determined by the
effective recombination rate to a given level~$i$, \change{since the
  bound-bound rates are always much faster}. In order to estimate the
contributions to the CMB spectral distortion by \ion{He}{ii}, we compute the
change in the population of level $i$ due to direct recombinations to that
level over a very short time interval $\Delta t$, i.e. $\Delta N_i\approx
N_{\rm e}\,N_{\rm c}\alpha^{\ion{He}{ii}}_{i}\,\Delta t$.

Because {\it all} the bound-bound transition rates in \ion{He}{ii} are 16
times larger than for hydrogen, the {\it relative} importance of the different
channels to lower states should remain the same as in hydrogen\footnote{Even
  the factors due to stimulated emission in the ambient blackbody radiation
  field are the same!}. Therefore one can assume that the relative number of
photons, $f_{ij}$, emitted in the transition $i\rightarrow j$ per additional
electron on the level $i$ is like for hydrogen at 4 times lower redshifts.

If we want to know how many of the emitted photons are observed in a fixed
frequency interval $\Delta\nu$ today ($z_{\rm obs}=0$) we also have to
consider that at higher redshift the expansion of the Universe is
faster. Hence the redshifting of photons through a given interval $\Delta \nu$
is accomplished in a shorter time interval. For a given transition, these are
related by $\Delta t= \frac{1+z}{H(z)\,\nu_{ij}}\,\Delta\nu$.
Then the change in the number of photons due to emission in the transition
$i\rightarrow j$ today should be proportional to 
\beal
\label{eq:DN_gamma_ij}
\Delta N_\gamma(\nu_{ij}) \sim 
\frac{f_{ij}(z_{\rm em})\,\Delta N_i(z_{\rm em})}{H(z_{\rm em})\,(1+z_{\rm em})^3}
\,\frac{(1+z_{\rm em})}{\nu_{ij}} \Delta\nu
\Abst{,}
\end{align}
where $z_{\rm em}$ is the redshift of emission, and the change of the volume
element due to the expansion of the Universe is taken into account by the
factor of $(1+z_{\rm em})^3$.
This now has to be compared with the corresponding change in the number of
photons emitted in the same transition by hydrogen, but at 4 times lower
emission redshift.

For hydrogenic atoms with change $Z$ the recombination rate, including
stimulated recombination within the ambient CMB blackbody, scales like 
\citep{Kaplan1970}
\beal
\label{eq:alphas_i}
\alpha_{i}
\propto \frac{Z^4}{T^{3/2}} \int_{h\nu_i/k T}^\infty \frac{{\rm d}x}{x^2}
\propto \frac{Z^2}{T^{1/2}}
\Abst{,}
\end{align}
where $\nu_i$ is the ionization frequency of the level $i$, $T$ is the
temperature of the plasma. It was assumed that $h\nu_i\ll k T$.
Therefore one finds
$\alpha^{\ion{He}{ii}}_{i}(4T)/\alpha^{\ion{H}{i}}_{i}(T)\sim 2$. Assuming
radiation domination one also has $H(z)/H(4z)\sim 1/16$.
Hence, we find $\Delta N_\gamma^{\ion{He}{ii}}(\nu_{ij}^\ion{He}{ii}, 4z_{\rm em})/\Delta
N_\gamma^{\ion{H}{i}}(\nu_{ij}^\ion{H}{i}, z_{\rm em})\sim 8\% \times 4^3 \, 2 / 16 \sim
64\%$.
Note that $(1+z_{\rm em})/\nu_{ij}^{\ion{H}{i}}\equiv (1+4z_{\rm
  em})/\nu_{ij}^{\ion{He}{ii}}$.
Prior to the epoch of $\ion{He}{iii}\rightarrow\ion{He}{ii}$ recombination the
release of photons by helium is amplified by a factor of $\sim 8$!

\subsection{Contributions due to \ion{He}{i}}
In the case of neutral helium, the highly excited levels are basically
hydrogenic. Therefore one does not expect any amplification of the emission
within loops prior to its recombination epoch.
Furthermore, the total period during which neutral helium can contribute
significantly is limited to the redshift range starting at the end of
$\ion{He}{iii}\rightarrow\ion{He}{ii}$ recombination, say $1600\lesssim z
\lesssim 6000$. Therefore, neutral helium typically will not be active over a
large range of redshifts.

Still there could be some interesting features appearing in connection with
the fine-structure transitions, which even within the standard computations
lead to strong negative features in the \ion{He}{i} recombination spectrum
\citep{Jose2007}. Also the spectrum of neutral helium, especially a high
frequencies, is more complicated than for hydrogenic atoms, so that some
non-trivial features might arise. We leave this problem for some future work,
and focus on the contributions of hydrogen and \ion{He}{ii}.

\section{Results for intrinsic $y$-type CMB distortions}
\label{sec:results}
Here we discuss the results for the changes in the recombination spectra of
hydrogen and \ion{He}{ii} for different values of the $y$-parameter.
Some of the computational details and the formulation of the problem
can be found in Appendix~\ref{app:comp}.
%

\begin{figure}
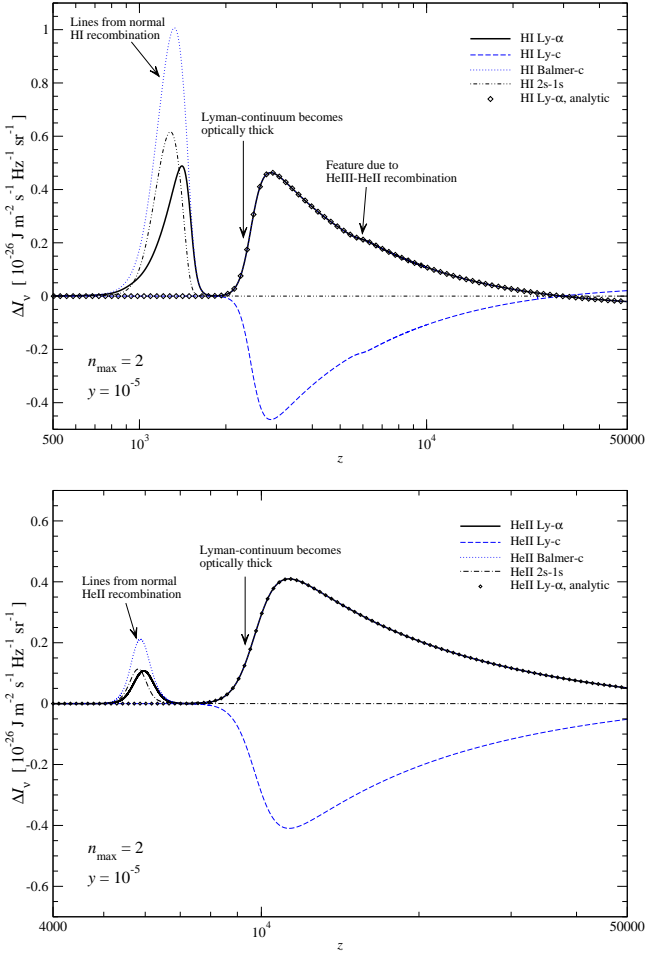

\centering 
\includegraphics[width=0.95\columnwidth]{./eps/Rates.2.y.diff_z.HI.eps}
\\[3mm]
\includegraphics[width=0.95\columnwidth]{./eps/Rates.2.y.diff_z.HeII.eps}
\caption{Spectral distortion, $\Delta I_\nu$, including 2 shells into our
  computations for different transitions as a function of redshift and for
  $y=10^{-5}$. All shown curves were computed using the $\delta$-function
  approximation for the intensity. The upper
  panel shows the results for hydrogen, the lower those for \ion{He}{ii}.
In both cases the analytic approximation for the Lyman-$\alpha$ line based on
Eq.~\eqref{eq:Sol_xi}, and including the escape probabilities in the
Lyman-$\alpha$ line and Lyman-continuum, are also shown.
}
\label{fig:Rates2}
\end{figure}
\subsection{The 2 shell atom}
\label{sec:2shellresults}
In order to understand the properties of the solution and also to check the
correctness of our computations we first considered the results including only
a small number of shells.
If we take 2 shells into account, we are only dealing with a few atomic
transitions, namely the Lyman- and Balmer-continuum, and the Lyman-$\alpha$
line. In addition, one expects that during the recombination epoch of the
considered atomic species (here \ion{H}{i} or \ion{He}{ii}) also the
2s-1s-two-photon decay channel will contribute, but very little before that
time.

In Fig.~\ref{fig:Rates2} we show the spectral distortion, $\Delta I_\nu$,
including 2 shells into our computations for different transitions as a
function of redshift\footnote{This is a convenient representation of the
  spectrum, when one is interested in the time-dependence of the photon
  release, rather than the observed spectral distortion in frequency space. To
  obtain the later, one simply has to plot the presented curves as a function
  of $\nu=\nu_{ij}/(1+z)$, where $\nu_{ij}$ is the restframe frequency of the
  considered transition.}.
It was assumed that energy was released in a single injection at $z_i=50000$,
leading to $y=10^{-5}$. All shown curves were computed using the
$\delta$-function approximation for the intensity \citep[for details
see][]{Jose2006}.
This approximation is not sufficient when one is interested in computing the
spectral distortions in frequency space.

Prior to the recombination epoch of the considered species one can find
pre-recombinational emission and absorption in the Lyman- and
Balmer-continuum, and the Lyman-$\alpha$ line, which would be
completely absent for $y=0$. As expected, during in the pre-recombinational
epochs the 2s-1s-two-photon transition is not important. This is because the
2s-1s transition is simply unable to compete with the $\sim 10^8$ times faster
Lyman-$\alpha$ transition while it is still optically thin.

\begin{figure}
\centering 
\includegraphics[width=0.95\columnwidth]{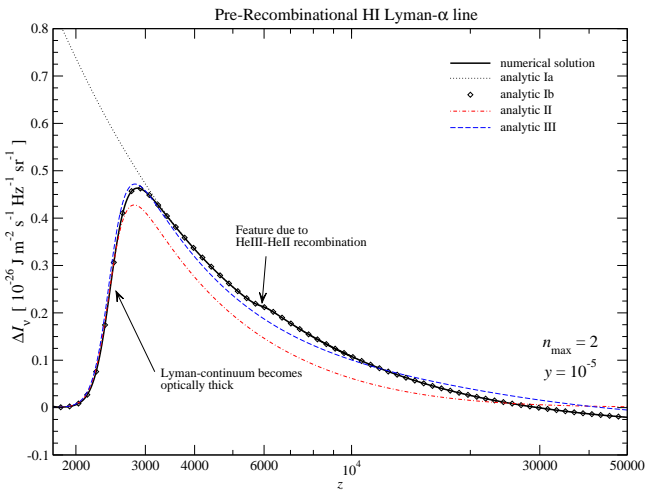}
\caption{Analytic representation of the pre-recombinational \ion{H}{i}
Lyman-$\alpha$ spectral distortions for the 2 shell atom and $y=10^{-5}$.
See text for explanations.
}
\label{fig:RatesAnalytic}
\end{figure}

Summing the spectral distortions due to the continuua, one finds cancellation
of the redshift-dependent emission at a level close to our numerical accuracy
(relative accuracy $\lesssim 10^{-4}$ for the spectrum). This is expected
because of electron number conservation: in the pre-recombinational epoch the
overall ionization state of the plasma is not affected significantly by the
small deviations of the background radiation from full equilibrium.
Therefore all electrons that enter an atomic species will leave it again, in
general via another route to the continuum.
This implies that $\sum_i \Delta R_{{\rm c}i}=0$, which is a general property
of the solution in the pre-recombinational epoch. 

If we look at the Lyman- and Balmer-continuum in the case of hydrogen we can
see that at redshifts $z\gtrsim 30000$ electrons are entering via the
Lyman-continuum, and leaving via the Balmer-continuum, while in the redshift
range $2000\lesssim z\lesssim 30000$ the opposite is true. As expected, for
$z\gtrsim 2000$ the Lyman-$\alpha$ transition closely follows the
Balmer-continuum, \change{since every electron that enters the 2p-state and
  then reaches the ground level, had to pass through the Lyman-$\alpha$
  transition}.
Using the analytic solution for the Lyman-$\alpha$ line as given in the
Appendix~\ref{app:analytic} we find excellent agreement with the numerical
results until the real recombination epoch is entered at $z\lesssim 2000$.

In the case of \ion{He}{ii} for the considered range of redshifts the
pre-recombinational emission ($z\gtrsim 7000$) is always generated in the loop
${\rm c}\rightarrow {\rm 2p} \rightarrow {\rm 1s}\rightarrow {\rm c}$. Again
we find excellent agreement with the analytic solution for the Lyman-$\alpha$
line. Note that for \ion{He}{ii} the total emission in the pre-recombinational
epoch is much larger than in the recombination epoch at $z\sim 6000$ (see
discussion in Sect.~\ref{sec:helium_cont}). 
The height of the maximum is even comparable with the \ion{H}{i}
Lyman-$\alpha$ line.

As one can see from Fig.~\ref{fig:Rates2}, at high redshifts all transitions
become weaker.
This is due to the fact that the restframe frequencies of all lines are in the
Rayleigh-Jeans part of the CMB spectrum, where the effective chemical
potential of the $y$-distortion (see Sect.~\ref{sec:ydist}) is dropping
like\footnote{Or more correctly $\mu(x)\approx 7.4\,x\,y$ if one also takes
into account the difference in the photon and electron temperature $\Te\approx
\Tg [1+5.4\,y]$ (see Appendix~\ref{app:problem}).}  $\mu(x)\approx 2x\,y$. 
This implies that at higher redshift all transitions are more and more within
a pure blackbody ambient radiation field.
On the other hand the effective chemical potential increases towards lower
redshift, so that also the strength of the transitions increases.
However, at $z\lesssim 3000$ in the case of hydrogen, and $z\lesssim 11000$
for \ion{He}{ii}, the escape probability in the Lyman-continuum (see
Appendix~\ref{app:problem} and Eq.~\eqref{eq:tau_con} for quantitative
estimates) starts to decrease significantly, so that the pre-recombinational
transitions cease.
The maximum in the pre-recombinational Lyman-$\alpha$ line is formed due to
this rather sharp transitions to the optically thick region in the
Lyman-continuum (see also Sect.~\ref{sec:Lya2shellresults} for more details).
%

\begin{figure}
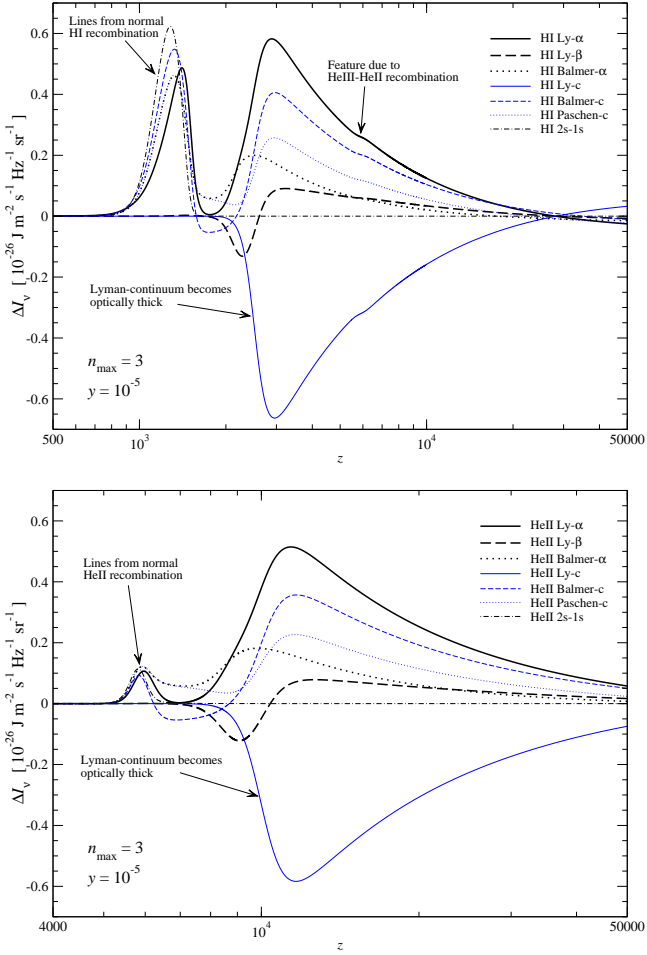

\centering 
\includegraphics[width=0.95\columnwidth]{./eps/Rates.3.y.diff_z.HI.eps}
\\[3mm]
\includegraphics[width=0.95\columnwidth]{./eps/Rates.3.y.diff_z.HeII.eps}
\caption{Spectral distortion, $\Delta I_\nu$, including 3 shells into our
  computations for different transitions as a function of redshift and for
  $y=10^{-5}$. For all shown curves we used the $\delta$-function
  approximation to compute the intensity. The upper panel shows the results
  for hydrogen, the lower those for \ion{He}{ii}.}
\label{fig:Rates3}
\end{figure}
\subsubsection{Analytic description of the pre-recombinational Lyman-$\alpha$ line}
\label{sec:Lya2shellresults}
One can understand the behavior of the solution for the spectral distortions
in more detail using our analytic description of the Lyman-$\alpha$ line as
given in Appendix~\ref{app:analytic}.

In Fig.~\ref{fig:RatesAnalytic} we show the comparison of different
approximations with the full numerical result.  
If we use the analytic approximation based on Eq.~\eqref{eq:Sol_xi}, but do
not include the escape probabilities in the \ion{H}{i} Lyman-$\alpha$ line and
\ion{H}{i} Lyman-continuum, then we obtain the dotted curve (quoted 'analytic
Ia' in the figure). The curve quoted 'analytic Ib' also includes the escape
probabilities as described in Appendix~\ref{app:Tauc}.
Comparing these curves shows that for the shape of the distortion at
$z\lesssim 3000$ the escape probabilities are very important. However,
although at this redshift the Sobolev optical depth in the \ion{H}{i}
Lyman-$\alpha$ line is roughly $14$ times larger than the optical depth in the
\ion{H}{i} Lyman-continuum, the derivation of Eq.~\eqref{eq:factor_a} shows
that the \ion{H}{i} Lyman-$\alpha$ escape probability only plays a secondary
role.

With the formulae in Appendix~\ref{app:moreapprox}, the spectral distortion
can be written in the form $\Delta I_\nu(z)=F(z)\times \Delta$.
If we use $F$ according to Eq.~\eqref{eq:factor_approx} and $\Delta\approx
\mu_{21}+\mu_{\rm 2pc}-\mu_{\rm 1sc}$, as derived in Eq.~\eqref{eq:imbal_b},
then we obtain the approximation quoted 'analytic II'. One can clearly see that
this approximation represents the global behavior, but is fails to explain the
Lyman-$\alpha$ absorption at $z\gtrsim 30000$. In fact within this
approximation the Lyman-$\alpha$ line should always be in emission, even at
very high redshifts, since there $\Delta\approx 3\,y\,x_{\rm 1sc}/32>0$.

If we also take into account higher order terms for the line imbalance
$\Delta$ according to Eq.~\eqref{eq:imbal_approx} then we obtain the curve
quoted 'analytic III', which is already very close to the full solution and
also reproduces the high redshift behavior, but starting at slightly higher
redshift ($z\sim 40000$ instead of $z\sim 30000$).
This is largely due to the approximations of the integrals
\eqref{eq:approx_sigav} over the photoionization cross-sections (in
particular $M_{-1}$). Still if one evaluates these integrals more accurately,
one does not recover the full solution, since the free-bound Gaunt-factors
were neglected. 

%
\subsection{The 3-shell atom}
\label{sec:3shellresults}
If one takes 3 shells into account, the situation becomes a bit more
complicated, since more loops connecting to the continuum are possible.
Looking at Fig.~\ref{fig:Rates3} again we find that the sum over all
transition in the continuua vanishes at redshifts prior to the actual
recombination epoch of the considered species.
At $z\lesssim 3000$ in the case of hydrogen, and $z\lesssim 11000$ for
\ion{He}{ii}, the escape probability in the Lyman-continuum becomes small.
For 2 shells this fact stopped the pre-recombinational emission until the
actual recombination epoch of the considered atomic species was entered (see
Fig.~\ref{fig:Rates2}).
However, for 3 shells electrons can now start to leave the 1s-level via the
Lyman-$\beta$ transition, and then reach the continuum through the
Balmer-continuum.
For both hydrogen and \ion{He}{ii} one can also see that the emission in the
Lyman-$\alpha$ line stops completely, once the Lyman-continuum is fully
blocked. In this situation only the loop ${\rm c}\rightarrow {\rm
  3}\rightarrow 2 \rightarrow {\rm c}$ via the Balmer-continuum is working.
Only when the main recombinational epoch of the considered species is entered,
the Lyman-$\alpha$ line is reactivated.

\begin{figure}
\centering 
\includegraphics[width=0.45\columnwidth]{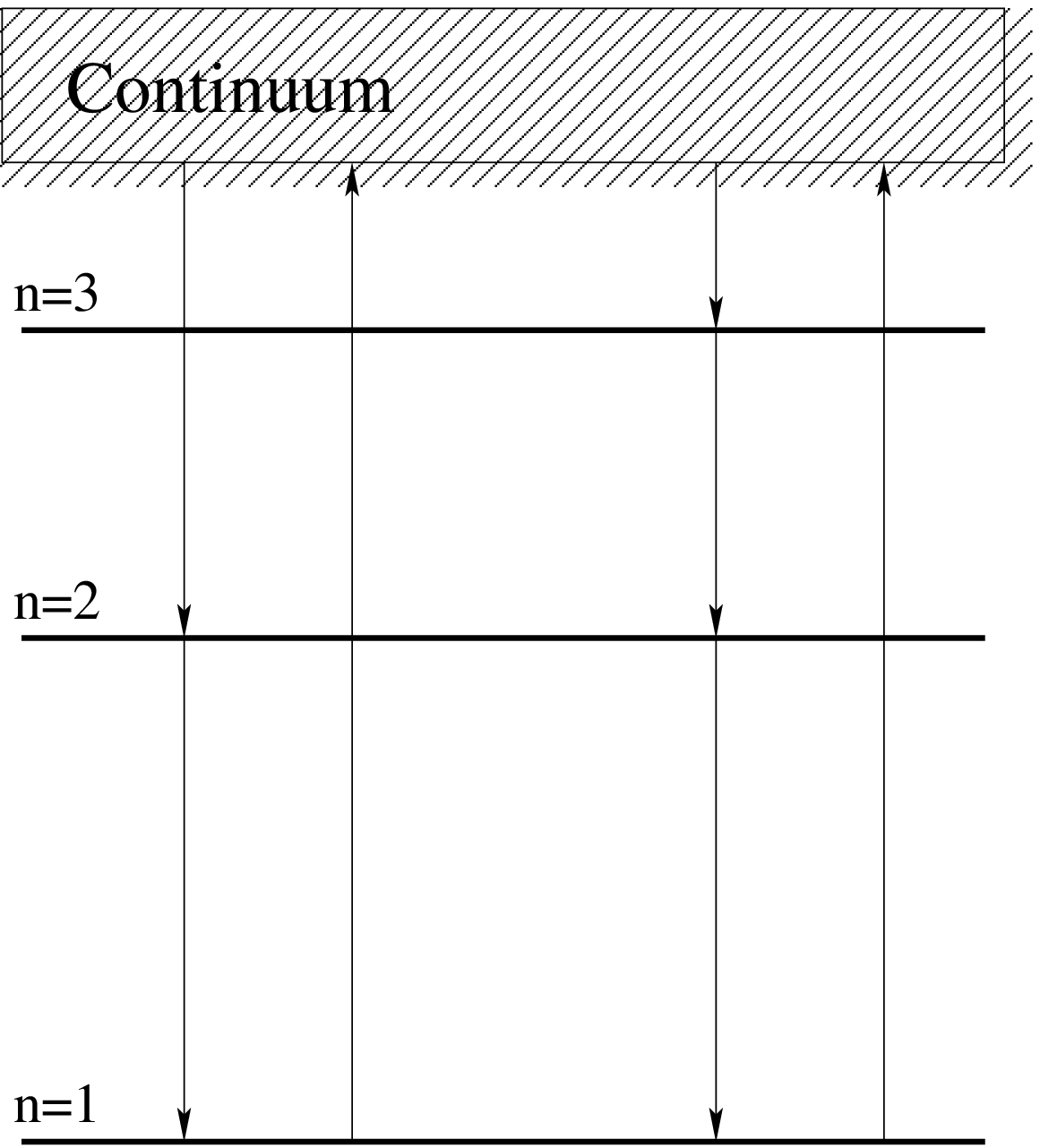}
\hspace{3mm}
\includegraphics[width=0.45\columnwidth]{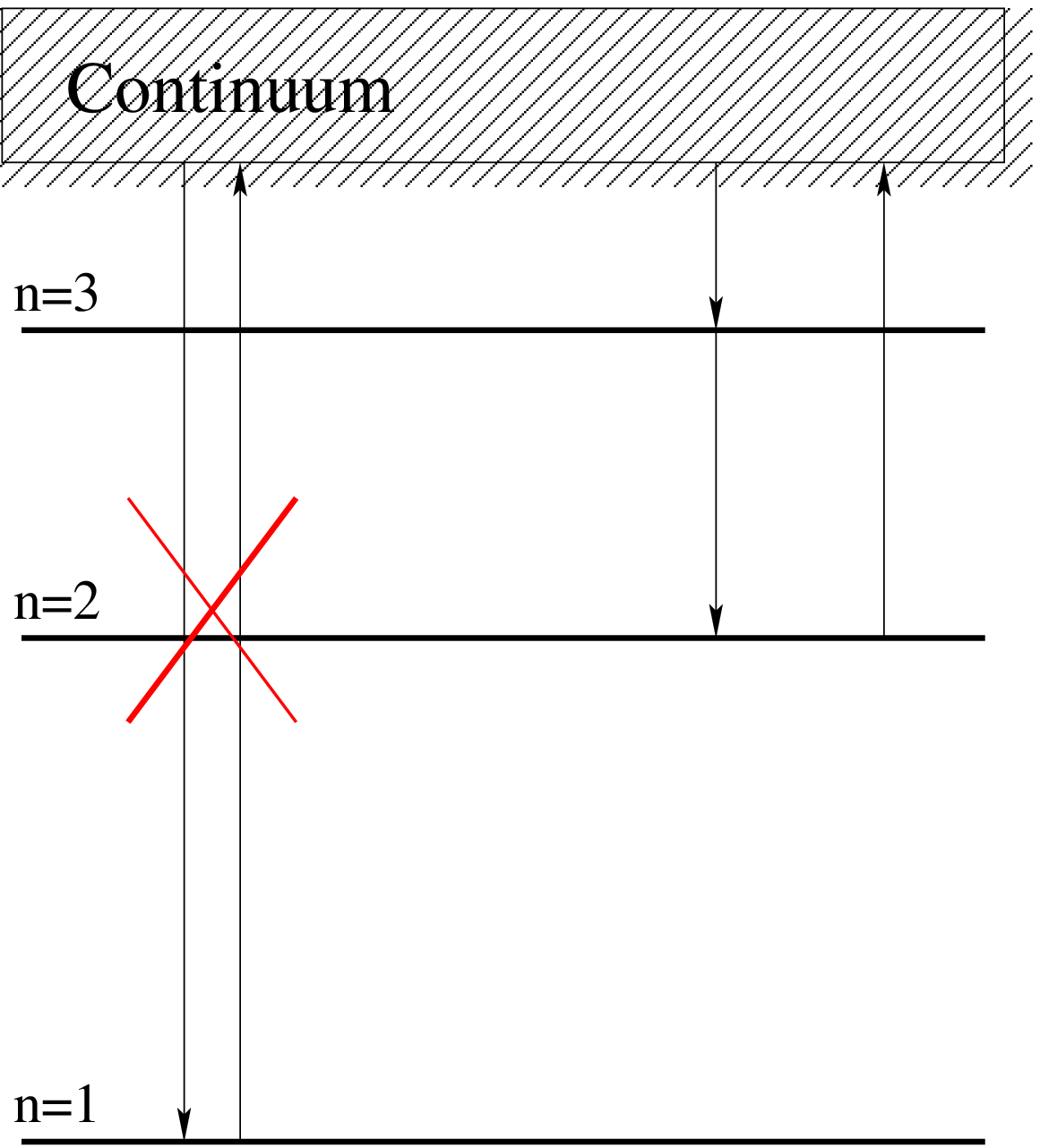}
\caption{Sketch of the main atomic loops for hydrogen and \ion{He}{ii} when
  including 3 shells. The left panel shows the loops for transitions that are
  terminating in the Lyman-continuum. The right panel shows the case, when the
  Lyman-continuum is completely blocked, and unbalanced transitions are
  terminating in the Balmer-continuum instead.}
\label{fig:loops}
\end{figure}
In Fig.~\ref{fig:loops} we sketch the main atomic loops in hydrogen and
\ion{He}{ii} when including 3 shells.
For $y=10^{-5}$, in the case of hydrogen the illustrated Lyman-continuum loops
work in the redshift range $2000\lesssim z \lesssim 30000$, while the
Balmer-continuum loop works for $1600\lesssim z \lesssim 2000$. In the case of
\ion{He}{ii} one finds $8000\lesssim z \lesssim \pot{1.2}{5}$ and
$6200\lesssim z \lesssim 8000$ for the Lyman- and Balmer-continuum loops,
respectively.
It is clear that in every closed loop one energetic photons is destroyed and
{\it at least} two photons are generated at lower frequencies. Including more
shells will open the possibility to generate more photons per loop, simply
because electrons can enter through highly excited levels and then
preferentially cascade down to the lowest shells via several intermediate
levels, leaving the atomic species taking the fastest available route back to
the continuum.
Below we will discuss this situation in more detail (see
Sect.~\ref{sec:Nloops}).

\begin{figure*}
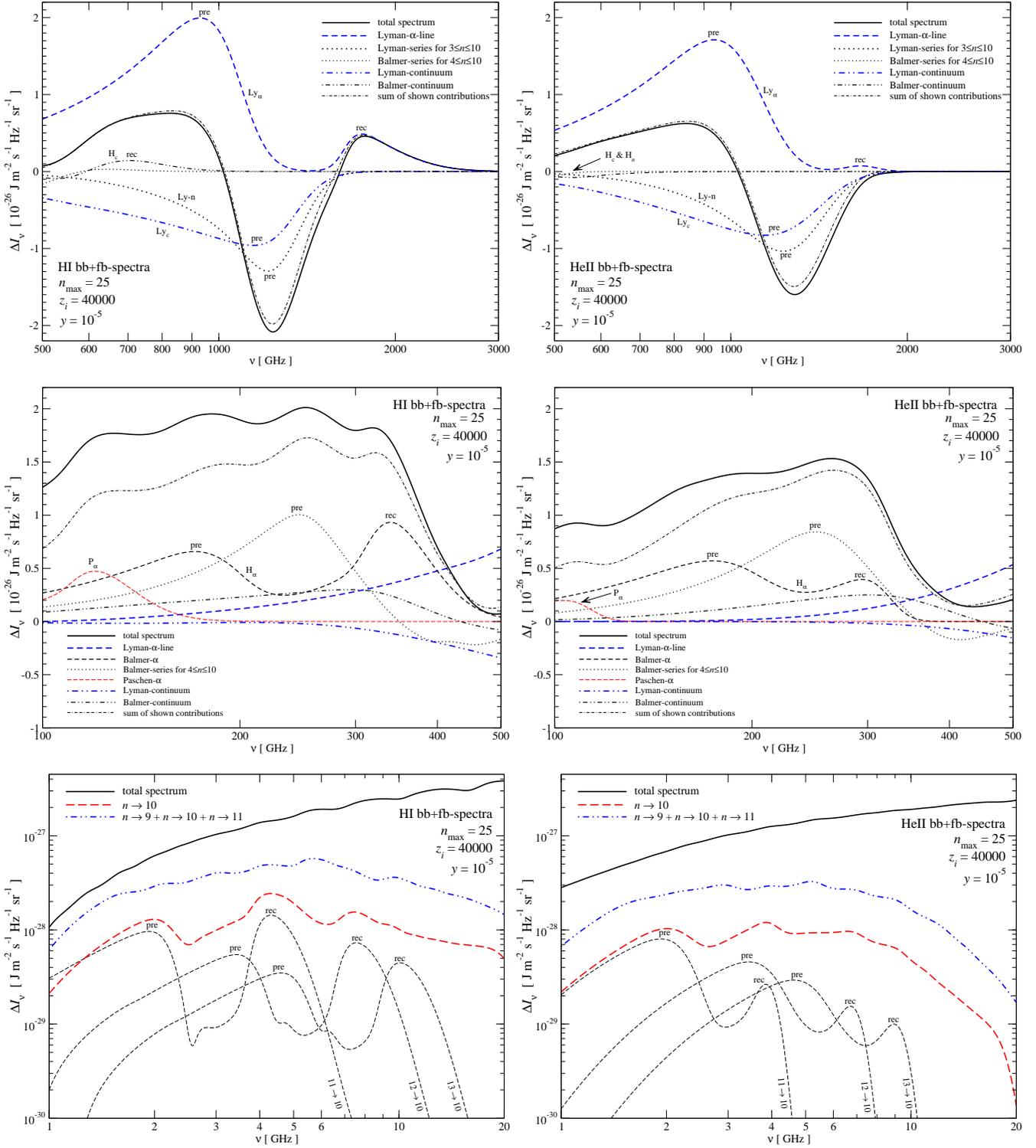

\centering 
\includegraphics[width=0.98\columnwidth]{./eps/contributions.25.y.very.high.eps}
\includegraphics[width=0.98\columnwidth]{./eps/contributions.HeII.25.y.very.high.eps}
\\[3mm]
\includegraphics[width=0.98\columnwidth]{./eps/contributions.25.y.high.eps}
\includegraphics[width=0.98\columnwidth]{./eps/contributions.HeII.25.y.high.eps}
\\[3mm]
\includegraphics[width=0.98\columnwidth]{./eps/contributions.25.y.low.eps}
\includegraphics[width=0.98\columnwidth]{./eps/contributions.HeII.25.y.low.eps}
\caption{Main contributions to the \ion{H}{i} (left panels) and \ion{He}{ii}
(right panels) spectral distortion at different frequencies for energy
injection at $z_i=40000$ and $y=10^{-5}$.
We have also marked those peaks coming (mainly) from the recombination epoch
('rec') and from the pre-recombination epoch ('pre') of the considered
atomic species.
%
}
\label{fig:HI.HeII.contributions.vh}
\end{figure*}

Figures~\ref{fig:Rates2} and \ref{fig:Rates3} both show that the
pre-recombinational lines are emitted in a typical redshift range $\Delta
z/z\sim 1$, while the signals from the considered recombinational epoch are
released within $\Delta z/z\sim 0.1-0.2$. For the pre-recombinational signal the
expected line-width is $\Delta\nu/\nu\sim 0.6-0.7$.
However, the overlap of several lines, especially at frequencies where
emission and absorption features nearly coincide, and the asymmetry of the
pre-recombinational line profiles, still leads to more narrow spectral
features with $\Delta\nu/\nu\sim 0.1-0.3$ (see Sect.~\ref{sec:25shellresults},
Fig.~\ref{fig:HI.HeII.contributions.vh}).
%
%

It is also important to mention that in all cases the actual recombination
epoch is not affected significantly by the small $y$-distortion in the ambient
photon field. There the deviations from Saha-equilibrium because of the
recombination dynamics dominate over those directly related to the spectral
distortion, and in particular the changes in the ionization history are tiny.

\begin{figure*}
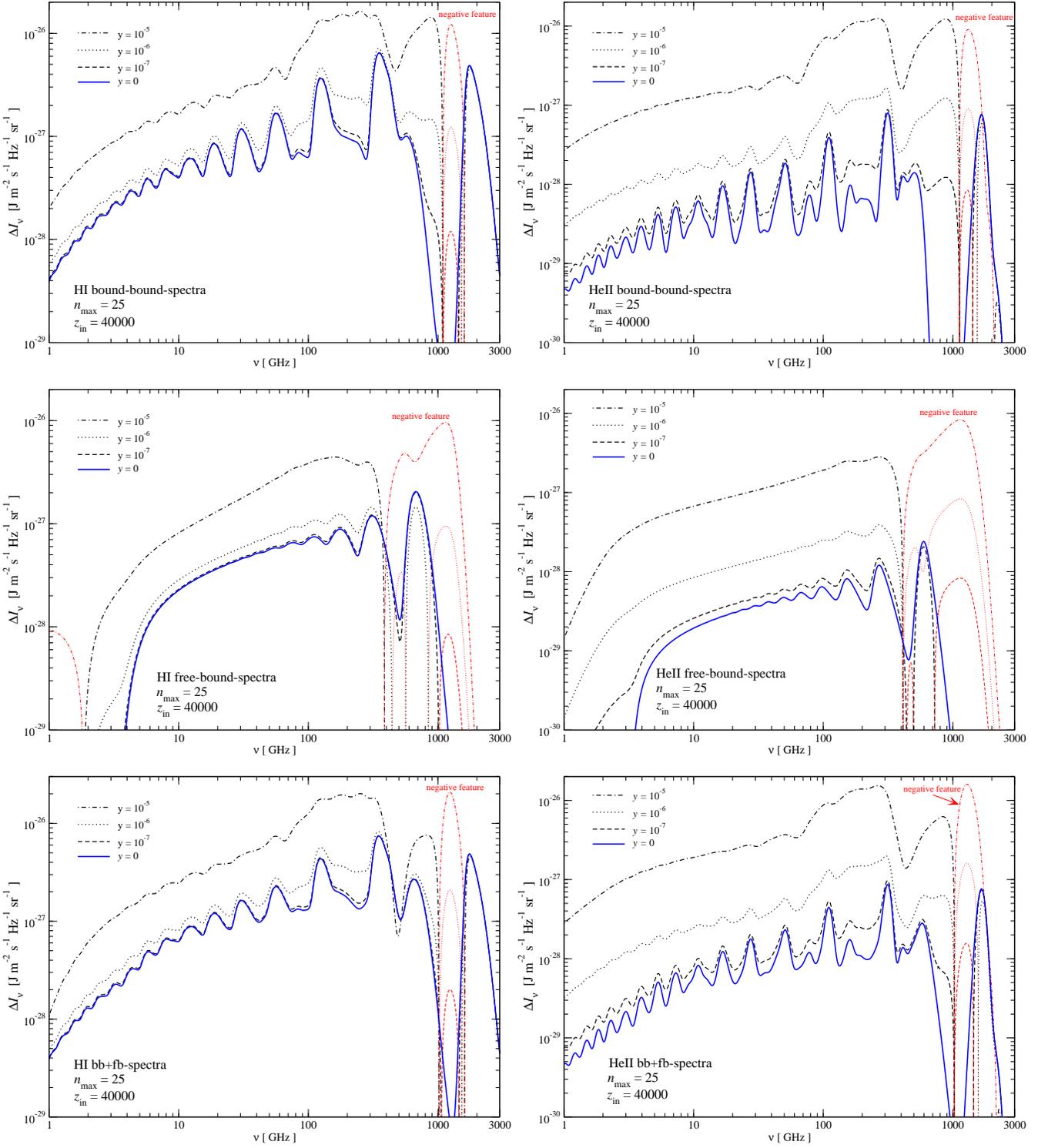

\centering 
\includegraphics[width=0.98\columnwidth]{./eps/spec.25.diff_y.HI.bb.eps}
\includegraphics[width=0.98\columnwidth]{./eps/spec.25.diff_y.HeII.bb.eps}
\\[3mm]
\includegraphics[width=0.98\columnwidth]{./eps/spec.25.diff_y.HI.fb.eps}
\includegraphics[width=0.98\columnwidth]{./eps/spec.25.diff_y.HeII.fb.eps}
\\[3mm]
\includegraphics[width=0.98\columnwidth]{./eps/spec.25.diff_y.HI.sum.eps}
\includegraphics[width=0.98\columnwidth]{./eps/spec.25.diff_y.HeII.sum.eps}
\caption{Contributions to the \ion{H}{i} (left panels) and \ion{He}{ii} (right
panels) recombination spectrum for different values of the initial
$y$-parameter. Energy injection was assumed to occur at $z_i=\pot{4}{4}$. In
each column the upper panel shows the bound-bound signal, the middle the
free-bound signal, and the lower panel the sum of both. The thin red lines
represent the overall negative parts of the signals.}
\label{fig:HI_HeII_y}
\end{figure*}

\subsection{The 25-shell atom}
\label{sec:25shellresults}
In this Section we discuss the results for our 25-shell computations. Given
the large amount of transitions, it is better to directly look at the spectral
distortion as a function of frequency. However, following the approach of
Sect~\ref{sec:2shellresults}, we have checked that the basic properties of the
first few lines and continuua as a function of redshift do not change
qualitatively in comparison to the previous cases. In particular, only the
Lyman- and Balmer-continuum become strongly negative, again in similar
redshifts ranges as for 2 or 3 shells.
In absorption the other continuua play no important role.

Note that for the spectral distortion now the impact of electron scattering
has to be considered and the free-bound contribution must be computed using
the full differential photoionization cross section (see
Appendix~\ref{app:elec}).

\subsubsection{Dependence of the distortion on the value of $y$}
\label{sec:ydep}
%
In Fig.~\ref{fig:HI_HeII_y} we show the contributions to the recombination spectrum
for different values of the initial $y$-parameter.
In addition, Fig.~\ref{fig:HI.HeII.contributions.vh} shows some of the main
contributions to the total hydrogen and \ion{He}{ii} spectral distortion in
more detail.  

\subsubsection*{Bound-bound transitions}
Focusing first on the contributions due to bound-bound transitions, one can
see that the standard recombination signal due to hydrogen is not very
strongly affected when $y=10^{-7}$, whereas the helium signal already changes
notably. Increasing the value of $y$ in both cases leads to an increase in the
overall amplitude of the distortion at low frequencies, and a large rise in
the emission and variability at $\nu\gtrsim 100\,$GHz. For \ion{H}{i} at low
frequencies the level of the signal changes roughly 5 times when increasing
the value of $y$ from $0$ to $10^{-5}$, while for \ion{He}{ii} the increase is
\change{even about a factor of $40$. This shows that in the
  pre-recombinational epoch \ion{He}{ii} indeed behaves similar to hydrogen,
  but with an amplification $\sim 8$ (see Sect.~\ref{sec:helium_cont}).}

At high frequencies, a strong {\it emission-absorption feature} appears in the
range $\nu\sim 500\,{\rm GHz}-1600\,$GHz, which is completely absent for
$y=0$. \change{With $y=10^{-5}$ from peak to peak this feature exceeds the
  normal Lyman-$\alpha$ distortion (close to $\nu\sim1750\,$GHz for \ion{H}{i}
  and $\nu\sim1680\,$GHz for \ion{He}{ii}) by a factor of $\sim 5$ for
  \ion{H}{i}, and about 30 times for \ion{He}{ii}. The absorption part is
  mainly due to the pre-recombinational Lyman-$\beta$, -$\gamma$,
  and~-$\delta$ transition, while the emission part is dominated by the
  pre-recombinational emission in the Lyman-$\alpha$ line (see also
  Fig.~\ref{fig:HI.HeII.contributions.vh} for more detail).}

It is important to note that in the case of \ion{He}{ii} most of the
recombinational Lyman-$\alpha$ emission ($\nu\sim1750\,$GHz) is completely
wiped out by the pre-recombinational absorption in the higher Lyman-series,
while for \ion{H}{i} only a very small part of the Lyman-$\alpha$ low
frequency wing is affected. This is possible only because the
pre-recombinational emission in the \ion{He}{ii} Lyman-series is so strongly
enhanced, as compared to the signal produced during the recombinational epoch.

\subsubsection*{Free-bound transitions}
Now looking at the free-bound contributions, one can again see that the
hydrogen signal changes much less with increasing value of $y$ than in the
case of \ion{He}{ii}. In both cases the variability of the free-bound signal
decreases at low frequencies, while at high frequencies a strong and broad
absorption feature appears, which is mainly due to the Lyman-continuum.
For $y=10^{-5}$ this absorption feature even completely erases the
Balmer-continuum contribution appearing during the actual recombination epoch
of the considered species.
It is 2 times stronger than the \ion{H}{i} Lyman-$\alpha$ line from the
recombination epoch, and in the case of \ion{He}{ii} it exceeds the normal
\ion{He}{ii} Lyman-$\alpha$ line by more than one order of magnitude.

However, except for the absorption feature at high frequencies the free-bound
contribution becomes practically featureless when reaching $y=10^{-5}$.
This is due to the strong overlap of different lines from the high redshift
part, since the characteristic width of the recombinational emission increases
like $\Delta\nu/\nu\sim k\Te/h \nu_{i\rm c}$ (see middle panels in
Fig.~\ref{fig:HI_HeII_y}).
In addition, the photons are released in a more broad range of redshifts (see
Sect.~\ref{sec:2shellresults} and \ref{sec:3shellresults}), also leading to a
lowering of the contrast of the spectral features from the recombinational
epoch.

\subsubsection*{Total distortion}
Also in the total spectra (see lower panels in Fig.~\ref{fig:HI_HeII_y}) one
can clearly see a strong absorption feature at high frequencies, which is
mainly associated with the Lyman-continuum and Lyman-series for $n>2$ (see
Fig.~\ref{fig:HI.HeII.contributions.vh} also).
For $y=10^{-5}$, in the case of hydrogen it exceeds the Lyman-$\alpha$ line
from recombination ($\nu\sim 1750\,$GHz) by a factor of $\sim 4$ at $\nu\sim
1250\,$GHz, while for \ion{He}{ii} it is even $\sim 20$ times stronger,
reaching $\sim 80\%$ of the corresponding hydrogen feature. Checking the level
of emission at low frequencies, as expected (see Sect.~\ref{sec:helium_cont}),
one can find that \ion{He}{ii} indeed contributes about $2/3$ to the total
level of emission.

As illustrated in the upper panels of Fig.~\ref{fig:HI.HeII.contributions.vh},
the emission-absorption feature at high frequencies is due to the overlap of
the pre-recombinational Lyman-$\alpha$ line (emission), and the combination of
the higher pre-recombinational Lyman-series and Lyman-continuum (absorption).
At intermediate frequencies (middle panels), the main spectral features are
due to the Balmer-$\alpha$, pre-recombinational Balmer-series from $n>3$ and
the Paschen-$\alpha$ transition, with some additional broad contributions to
the overall amplitude of the bump coming from higher continuua.

The lower panels of Fig.~\ref{fig:HI.HeII.contributions.vh} show, the separate
contributions to the bound-bound series for the 10th shell as an example.
One can notice that in the case of hydrogen the recombinational and
pre-recombinational emission have similar amplitude, while for \ion{He}{ii}
the pre-recombinational signal is more than one order of magnitude larger (see
Sect.~\ref{sec:helium_cont} for explanation). In both cases the
pre-recombinational emission is much broader than the recombinational signal,
again mainly due to the time-dependence of the photon emission process (see
Sect.~\ref{sec:2shellresults} and \ref{sec:3shellresults}), but to some extent
also because of electron scattering.
%
%

\begin{figure}
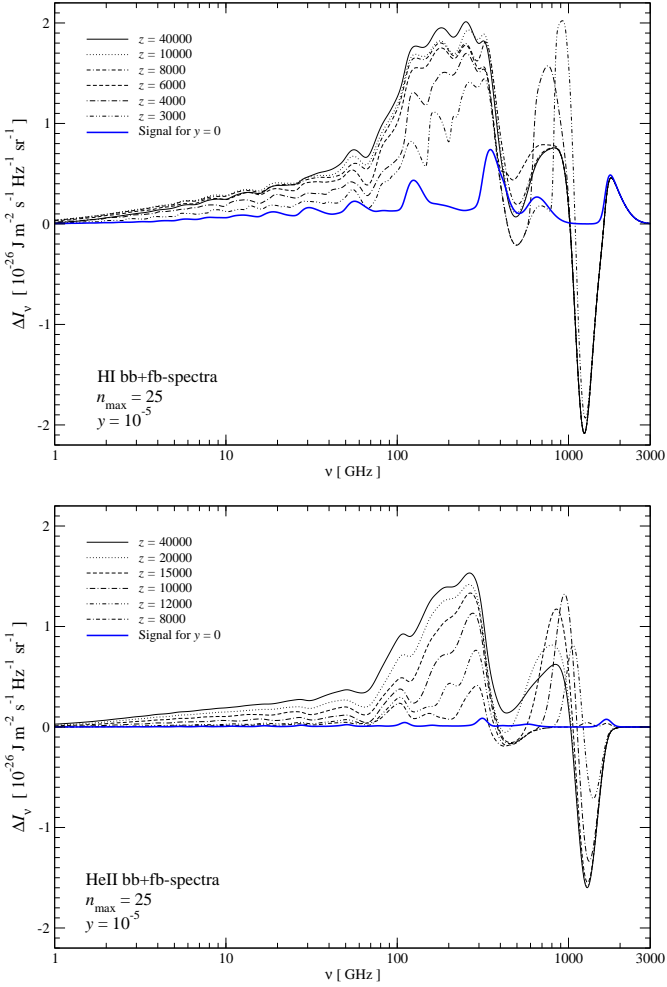

\centering 
\includegraphics[width=0.98\columnwidth]{./eps/spec.25.y.diff_z.HI.sum.lin.3.eps}
\\[3mm]
\includegraphics[width=0.98\columnwidth]{./eps/spec.25.y.diff_z.HeII.sum.lin.3.eps}
\caption{\ion{H}{i} (upper panel) and \ion{He}{ii} (lower panel) recombination
spectra for different energy injection redshifts.}
\label{fig:HI_HeII.diff_z}
\end{figure}

\begin{figure}
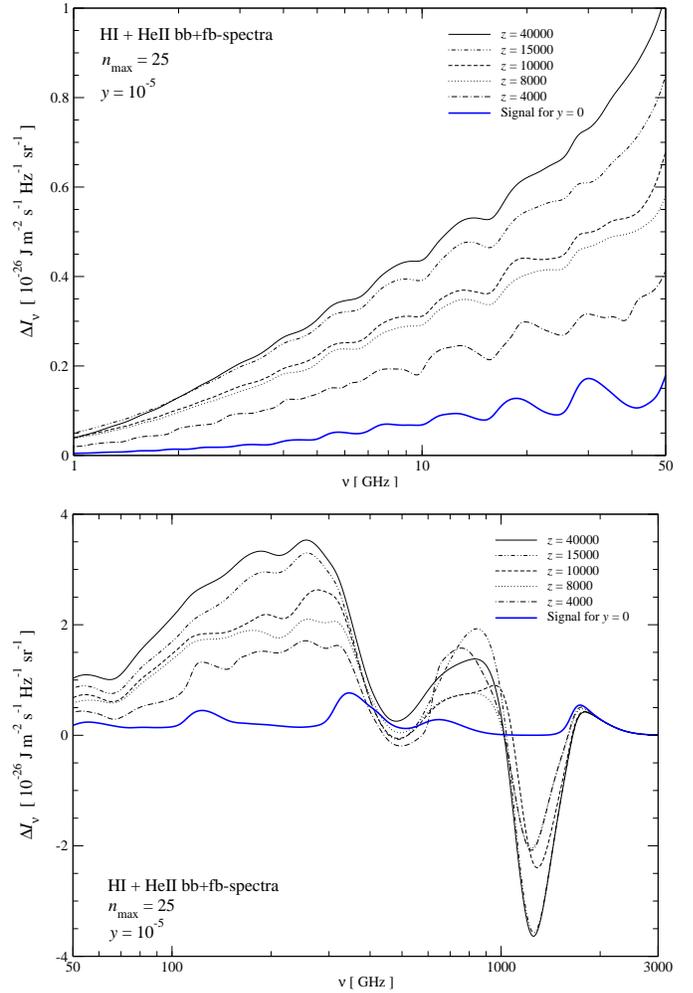

\centering 
\includegraphics[width=0.98\columnwidth]{./eps/HI.HeII.diff_y.low.eps}
\\[2mm]
\includegraphics[width=0.98\columnwidth]{./eps/HI.HeII.diff_y.high.eps}
\caption{Total \ion{H}{i} + \ion{He}{ii} recombination spectra for different
energy injection redshifts. The upper panel shows details of the spectrum at
low, the lower at high frequencies.}
\label{fig:HI.HeII.diff_z.hl}
\end{figure}



\begin{table}
\caption{Approximate number of photons and loops per nucleus for
$z_i=\pot{4}{4}$ and different values of $y$.}
\label{tab:one}
\centering
{
\begin{tabular}{@{}crrrr}
\hline
\hline
             & $y=0$ & $y=10^{-7}$ & $y=10^{-6}$ & $y=10^{-5}$ \\
\hline
$N^{\ion{H}{i}}_{\rm Ly-c}$ 
& $0\times N_{\rm H}$ & $-0.028\times N_{\rm H}$ & $-0.28\times N_{\rm H}$ & $-2.74\times N_{\rm H}$ 
\\
$N^{\ion{H}{i}}_{\rm Ly-\alpha}$ 
& $0.42\times N_{\rm H}$ & $0.47\times N_{\rm H}$ & $0.92\times N_{\rm H}$ & $5.37\times N_{\rm H}$ 
\\
$N^{\ion{H}{i}}_{\rm bb}$ 
& $2.49\times N_{\rm H}$ & $2.59\times N_{\rm H}$ & $3.45\times N_{\rm H}$ & $12.04\times N_{\rm H}$ 
\\
\hline
$\NloopH$ 
& $0\times N_{\rm H}$ & $0.032\times N_{\rm H}$ & $0.32\times N_{\rm H}$ & $3.24\times N_{\rm H}$ 
\\
\hline
\hline
$N^{\ion{He}{ii}}_{\rm Ly-c}$ 
& $0\times N_{\rm He}$ & $-0.27\times N_{\rm He}$ & $-2.67\times N_{\rm He}$ & $-26.6\times N_{\rm He}$ 
\\
$N^{\ion{He}{ii}}_{\rm Ly-\alpha}$ 
& $0.55\times N_{\rm He}$ & $1.06\times N_{\rm He}$ & $5.68\times N_{\rm He}$ & $51.8\times N_{\rm He}$ 
\\
$N^{\ion{He}{ii}}_{\rm bb}$ 
& $2.48\times N_{\rm He}$ & $3.49\times N_{\rm He}$ & $12.5\times N_{\rm He}$ & $103\times N_{\rm He}$ 
\\
\hline
$\NloopHe$ 
& $0\times N_{\rm He}$ & $0.28\times N_{\rm He}$ & $2.91\times N_{\rm He}$ & $30.1\times N_{\rm He}$ 
\\
\hline
\hline
\end{tabular}
}
\end{table}
\subsubsection*{Number of photons and loops}
\label{sec:Nloops}
Using the free-bound spectrum, one can also compute the total number of loops,
$\Nloop$, that were involved into the production of photons. This is possible,
since only during the recombination epoch electrons will terminate in the
1s-state. Therefore the total number of photons emitted in the free-bound
spectrum is very close to $\sim 1\gamma$ per nucleus, while when looking at
the positive or negative part of this contribution one should find $\sim
(\Nloop+1)\gamma$ and $\sim -\Nloop\gamma$ per nucleus, respectively.

In Table~\ref{tab:one} we give a few examples, also comparing with the number
of photons emitted for $y=0$. 
One can see that the number of loops per nucleus scales roughly proportional
to the values of $y$, i.e. $\NloopH\sim 3.2\times [y/10^{-5}]$ and
$\NloopHe\sim 30\times [y/10^{-5}]$. If one would consider a lower injection
redshift the proportionality constant should decrease.
Also when including more shells $\Nloop$ should become larger.
Furthermore, the number of loops per nucleus is about one order of magnitude
larger for \ion{He}{ii} than for hydrogen. As explained in
Sect.~\ref{sec:helium_cont}, this is due to the amplification of transitions
in the case of hydrogenic helium at high redshifts.  
Comparing the number of photons absorbed in the Lyman-continuum, one can see
that in practically all cases $\sim 90\%$ of all loops are ending there.

If we take the total number of photons per nucleus emitted in the bound-bound
transitions and subtract the number of photons emitted for $y=0$, we can
estimate the loop-efficiency, $\epsilon_{\rm loop}$, or number of bound-bound
photons generated per loop prior to the recombination epoch. For hydrogen one
finds $\epsilon_{\rm loop}\sim 2.9-3.1$, while for \ion{He}{ii} one has
$\epsilon_{\rm loop}\sim 3.3-3.6$.
Similarly one obtains an loop efficiency of $\epsilon_{\rm loop}\sim 1.7-1.8$
for both the \ion{H}{i} and \ion{He}{ii} Lyman-$\alpha$ lines.
As expected these numbers are rather independent of the value of $y$, since
they should reflect an atomic property. They should also be rather independent
of the injection redshift, which mainly affects the total number of loops and
thereby the total number of emitted photons.
However, the loop efficiency should still increase when including more shells
in the computation.
%

\subsubsection{Dependence of the distortion on the redshift of energy injection}
\label{sec:zdep}
%
To understand how the pre-recombinational emission depends on the redshift at
which the energy was released, in Fig.~\ref{fig:HI_HeII.diff_z} we show a
compilation of different cases for the total \ion{H}{i} and \ion{He}{ii}
signal.
In Fig.~\ref{fig:HI.HeII.diff_z.hl} we also present the sum of both in more
detail.

\begin{figure*}
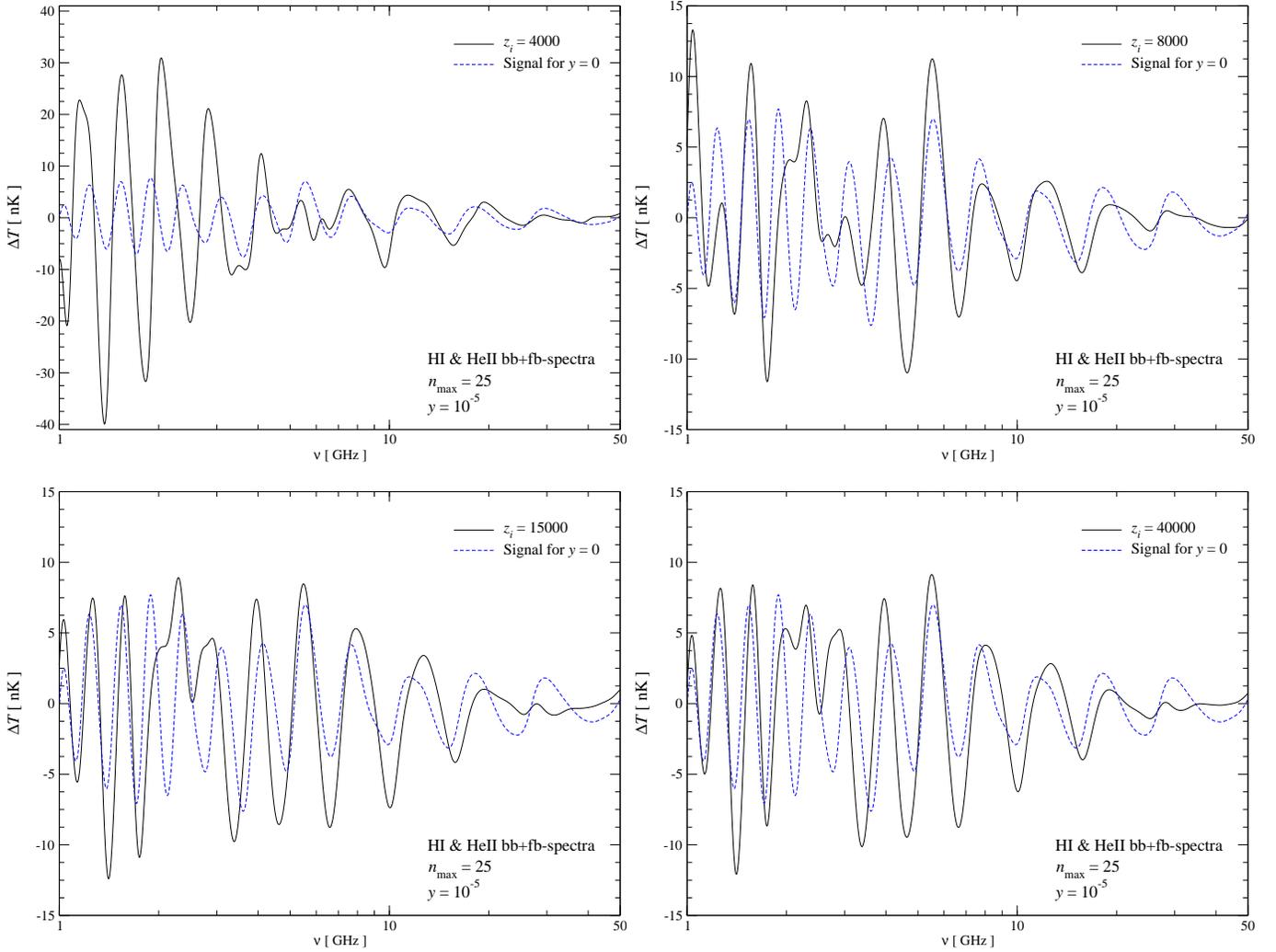

\centering 
\includegraphics[width=\columnwidth]{./eps/DT.n_25.1GHz.z_4000.eps}
\includegraphics[width=\columnwidth]{./eps/DT.n_25.1GHz.z_8000.eps}
\\[3mm]
\includegraphics[width=\columnwidth]{./eps/DT.n_25.1GHz.z_15000.eps}
\includegraphics[width=\columnwidth]{./eps/DT.n_25.1GHz.z_40000.eps}
\caption{Comparison of the variable component in the \ion{H}{i} + \ion{He}{ii}
  bound-bound and free-bound recombination spectra for single energy injection
  at different redshifts. In all cases the computations were performed
  including 25 shells and $y=10^{-5}$. The blue dashed curve in all panel
  shows the variability in the normal \ion{H}{i} + \ion{He}{ii} recombination
  spectrum (equivalent to energy injection below $z\sim 800$ or $y=0$) for
  comparison.}
\label{fig:DT.25.diff_z}
\end{figure*}
\subsubsection*{Features at high frequencies}
For all shown cases the absolute changes in the curves are strongest at high
frequencies ($\nu\gtrsim 100\,$GHz). Generally speaking, one can
again find a rather broad bump at $100\,{\rm GHz}\lesssim\nu\lesssim
400\,{\rm GHz}$, followed by an emission-absorption feature in the frequency
range $500\,{\rm GHz}\lesssim\nu\lesssim 1600\,{\rm GHz}$. In particular the
strength and position of this emission-absorption feature depends strongly on
the redshift of energy injection.

Regarding the broad high-frequency signature, it is more important that the
variability is changing, rather than the increase in the overall amplitude.
For example, in the frequency range $100\,{\rm GHz}\lesssim\nu\lesssim
400\,{\rm GHz}$ the normal recombinational signal has $\sim 2$ spectral
features, while for injection at $z_i\sim 4000$ roughly 4 features are
visible, which in this case only come from hydrogen, since at $z\sim 4000$
\ion{He}{ii} is already completely recombined. 
Note that also neutral helium should add some signal, which was not included
here. Nevertheless, we expect that this contribution is not strongly amplified
like in the case of \ion{He}{ii} (see Sect.~\ref{sec:helium_cont}), and hence
should not add more than $10-20\%$ to the total signal.

\subsubsection*{Variability at low frequencies}
Focusing on the spectral distortions at low frequencies, the overall level of
the distortion in general increases for higher redshifts of energy release.
However, there is also some change in the {\it variability} of the spectral
distortions.
In order to study this variability in more detail, in each case we performed a
smooth spline fit of the total \ion{H}{i} + \ion{He}{ii} recombination
spectrum and then subtracted this curve from the total spectrum. The
remaining modulation of the CMB intensity was then converted into variations
of the CMB brightness temperature using the relation $\Delta T/T_0 = \Delta I
/ B_{\nu}$, where $B_\nu$ is the blackbody intensity with temperature
$T_0=2.725\,$K.

Figure~\ref{fig:DT.25.diff_z} shows the results of this procedure for several
cases.
It is most striking that the amplitude of the variable component decreases
with increasing energy injection redshift. 
This can be understood as follow:
we have seen in Sect.~\ref{sec:2shellresults} and \ref{sec:3shellresults} that
for very early energy injection most of the pre-recombinational emission is
expected to arise at $z\sim 3000$ for hydrogen, and $z\sim 11000$ for helium,
\change{(i.e. the redshifts at which the Lyman-continuum of the considered
atomic species becomes optically thick)} with a typical line-width
$\Delta\nu/\nu\sim 1$.
In this case the total variability of the signal is mainly due to the
non-trivial superposition of many broad neighboring spectral features.
Most importantly, very little variability will be added by the high redshift
wing of the pre-recombinational lines and in particular the {\it beginning} of
the injection process. This is because (i) at high $z$ the emission is much
smaller (cf. Fig.~\ref{fig:Rates2} and \ref{fig:Rates3}), and (ii) electron
scattering broadens lines significantly, smoothing any broad feature even more
(see Appendix~\ref{app:elec}).

On the other hand, when the energy injection occurs at lower redshift this
increases the variability of the signal because (i) electron scattering in the
case of single momentary energy release does not smooth the step-like feature
due to the beginning of the injection process as strongly, and (ii) the total
emission amplitude and hence the step-like feature increases (see
Fig~\ref{fig:Rates2} and \ref{fig:Rates3}).
Once the injection occurs very close to the redshift at which the
Lyman-continuum is becoming optically thick (see
Sect.~\ref{sec:2shellresults}), i.e. where the pre-recombinational emission
has an extremum, this therefore should lead to a strong increase of the
variability.
On the other hand, for energy injection well before this epoch the atomic
transitions lead to an increase in the overall amplitude of the distortions
rather than the variability.

Indeed this can be also seen in Fig.~\ref{fig:DT.25.diff_z}, where for
$z_i=4000$ the variable component is $\sim 3-8$ times larger than the
normal recombinational signal, with a peak to peak amplitude of $\sim
50-70\,$nK instead of $\sim 10-15\,$nK at frequencies around $\sim 1.5\,$GHz.
Even for $z_i\gtrsim 15000$ the amplitude of the variable component is still
$1.5-2$ times larger than in the case of standard recombination, but it
practically does not change anymore when going to higher injection redshifts.
For $z_i\sim 11000$ one expects a similarly strong increase in the variability
as for $z_i\sim 4000$, but this time due to \ion{He}{ii}.
In addition to the change in amplitude of the variable component, in all cases
the signal is shifted with respect to the normal recombinational signal.
\change{These shifts should also make it easier to distinguish the signatures
  from pre-recombinational energy release from those arising because of normal
  recombination}.
%

\begin{figure}
\centering 
\includegraphics[width=\columnwidth]{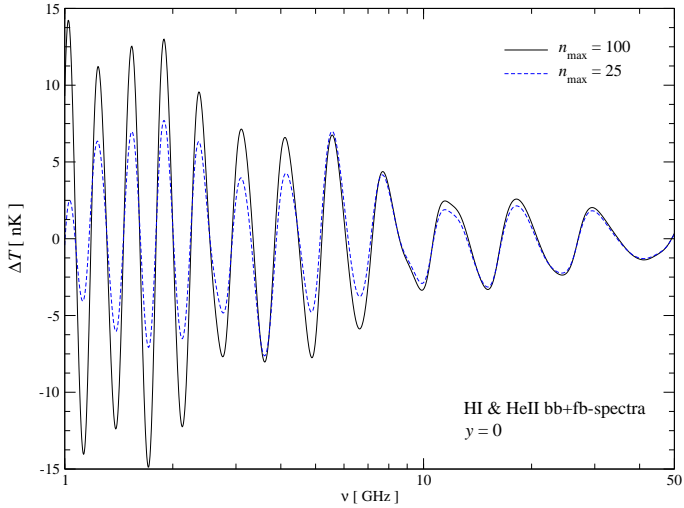}
\caption{Comparison of the variable component in the standard ($y=0$)
\ion{H}{i} + \ion{He}{ii} bound-bound and free-bound recombination spectrum
for $n_{\rm max}=100$ and 25.}
\label{fig:DT.100.25}
\end{figure}

%
It is important to mention that the total amplitude of the variable component,
should still increase when including more shells into the computation.
As shown in \citet{Chluba2007}, for $y=0$ in particular the overall level of
recombinational emission at low frequencies strongly depends on the
completeness of the atomic model. Similarly, the variability of the
recombination spectrum changes.
We illustrate this fact in Fig.~\ref{fig:DT.100.25}, where we compare the
variability in the \ion{H}{i} + \ion{He}{ii} recombination spectrum for 25
shells ($y=0$), with the one obtained in our 100 shell computations
\citep{Chluba2007, Jose2007}.
As one can see, at low frequencies ($\nu\sim 1-3\,$GHz) the amplitude of the
variable component increases by more than a factor of 2 when including 100
shells, \change{reaching a peak to peak amplitude of $\sim 30\,$nK}. This is
due to the fact that for a more complete atomic model additional electrons are
able to pass through a particular transition between highly excited states.

\begin{figure}
\centering 
\includegraphics[width=\columnwidth]{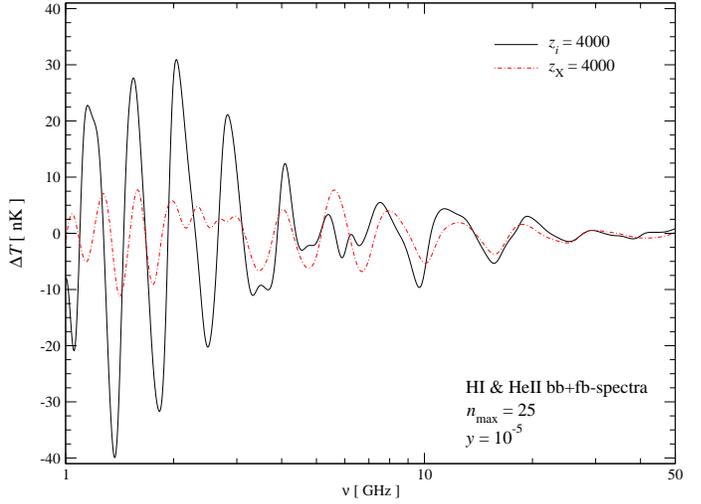}
\\[3mm]
\includegraphics[width=\columnwidth]{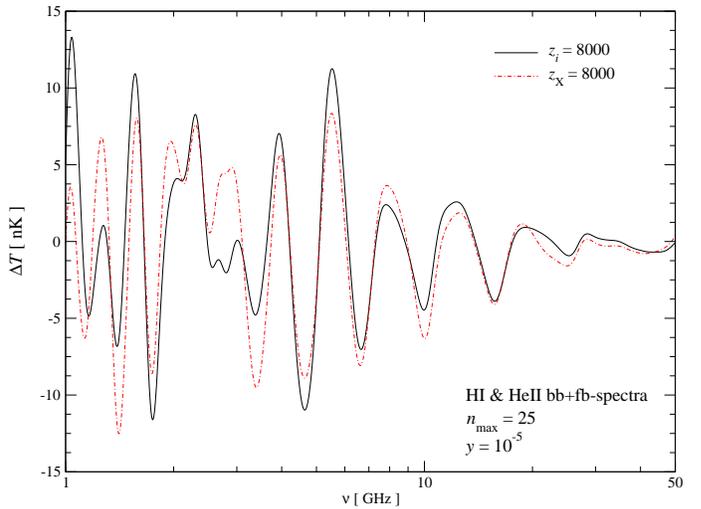}
%
\caption{Comparison of the variable component in the \ion{H}{i} + \ion{He}{ii}
bound-bound and free-bound recombination spectra for single energy injection
(black solid curves) and energy injection due to long-lived decaying particles
with different lifetimes (red dashed-dotted curves).
In all cases the computations were performed including 25 shells and a maximal
$y$-parameter $y=10^{-5}$. }
\label{fig:DT.25.diff_z.decay}
\end{figure}
\subsubsection{Dependence of the distortion on the energy injection history}
\label{sec:yzdep}
%
Until now we have only considered cases for single momentary energy injection.
However, physically this may not be very realistic, since most of the possible
injection mechanisms release energy over a more broad range of redshifts.
Also the discussion in Sect.~\ref{sec:zdep} has shown that for single
injection a large part of the variability can be attributed to the inset of
the energy release. Therefore it is important to investigate the potential
signatures of other injection mechanisms.

For the signals under discussion long lived decaying particles are most
interesting. In Sect.~\ref{sec:ydist_decay} we have given some simplified
analytic description of this problem.
In Fig.~\ref{fig:DT.25.diff_z.decay} we show the variable component for the
CMB spectral distortion due to the presence of hydrogen and \ion{He}{ii} at
low frequencies, in the case of single injection and for energy release due to
long-lived decaying particles.
It is clear that for $z_{\rm X}=4000$ the variability is
significantly smaller than in the case of single energy
injection. This is due to the fact that the inset of the atomic transitions is
much more gradual than in the case of single injection. However one should
mention that for energy injection due to decaying particles the effective
$y$-parameter at $z=4000$ is still only $\sim 1/3$ of the maximal value, so
that the level of variability cannot be directly compared with the case of
single energy injection.
Nevertheless, the structure of the variable component still depends
non-trivially on the effective decay redshift, so that one in principle should
be able to distinguish different injection scenarios.

Similarly, one could consider the case of annihilating particles. However,
here energy is effectively released at higher redshifts\footnote{This
  conclusion depends also on the temperature/energy dependence of the
  annihilation cross section. We assumed $s$-wave annihilation \citep[e.g.
  see][]{McDonald2001}.}  and also within a much broader redshift interval.
In this case, one has to follow the evolution of the CMB spectrum due to this
heating mechanism from an initial $\mu$-type distortion to a partial $y$-type
distortion in more detail.  
Also one can expect that the redistribution of photons via electron scattering
will become much more important (see Appendix~\ref{app:elec}), and that the
free-free process will strongly alter the number of photons emitted via atomic
transitions (see Appendix~\ref{app:ff_est}).
In addition, collisional processes may become significant, in particular those
leading to transitions among different bound-bound level, or to the continuum,
since they are not associated with the emission of photons.
This problem will be left for some future work.  
%

\section{Discussion and conclusions}
\label{sec:disc}
\label{sec:conc}
\begin{figure}
\centering 
\includegraphics[width=\columnwidth]{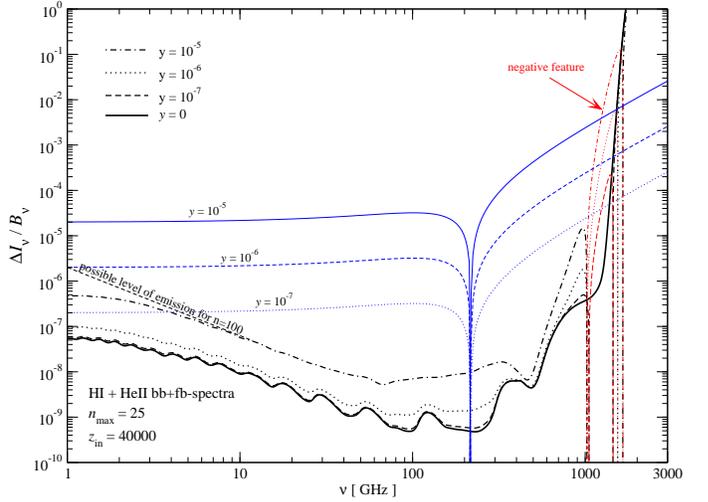}
\caption{Spectral distortions relative to the CMB blackbody spectrum,
  $B_\nu$. The thin blue curves show the absolute value of a $y$-type
  distortion. At low frequencies we indicate the expected level of emission
  when including more shells in our computations.}
\label{fig:comp_w_y}
\end{figure}

In the previous Sections we have shown in detail how intrinsic $y$-type CMB
spectral distortions modify the radiation released due to atomic
transitions in primordial hydrogen and \ion{He}{ii} at high redshifts.
We presented the results of numerical computations including 25 atomic shells
for both \ion{H}{i} and \ion{He}{ii}, and discussed the contributions of several
individual transitions in detail (e.g. see
Fig.~\ref{fig:HI.HeII.contributions.vh}), also taking the broadening of lines
due to electron scattering into account.
As examples, we investigate the case of instantaneous energy release
(Sect.~\ref{sec:zdep}) and exponential energy release (Sect.~\ref{sec:yzdep})
due to long lived decaying particles, separately.
Our computations show that several additional photons are released during the
pre-recombinational epoch, which in terms of number can strongly exceed those
from the recombinational epoch (see Sect.~\ref{sec:Nloops}). 
The number of loops per nucleus scales roughly proportional to the values of
$y$, i.e. $\NloopH\sim 3.2\times [y/10^{-5}]$ and $\NloopHe\sim 30\times
[y/10^{-5}]$ for hydrogen and \ion{He}{ii} respectively, where effectively
about 3 photons per loop are emitted in the bound-bound transition.
Due to the non-trivial overlap of broad neighboring pre-recombinational lines
(from bound-bound and free-bound transitions), rather narrow
($\Delta\nu/\nu\sim 0.1-0.3$) spectral features on top of a broad continuum
appear, which both in shape and amplitude depend on the time-dependence of the
energy injection process and the value of the intrinsic $y$-type CMB
distortion.
At high frequencies ($\nu\sim 500\,{\rm GHz}-1600\,$GHz) an
emission-absorption feature is formed, which is completely absent for $y=0$,
and is mainly due to the superposition of pre-recombinational emission in the
Lyman-$\alpha$ line, and the higher Lyman-series and Lyman continuum.
Looking at Fig.~\ref{fig:comp_w_y} it becomes clear that this absorption
feature (close to $\nu\sim 1400\,$GHz) in all shown cases even exceeds the
intrinsic $y$-distortion. For $y=10^{-5}$ it even reaches $\sim 10\%$ of the
CMB blackbody intensity.
Unfortunately, it appears in the far Wien-tail of the CMB spectrum, where
especially the cosmic infrared background due to \change{dusty star-forming
galaxies is dominant \citep{Fixsen1998, Lagache2005}.}
Still one may hope to be able to extract such spectral feature in the future.

%
One should stress that still {\it all} the discussed additional
pre-recombinational spectral distortions are in general {\it small} in
comparison with the intrinsic $y$-distortion.
As Fig.~\ref{fig:comp_w_y} shows, the amplitude of the additional distortions
is typically well below 1\% of the CMB y-distortion.
However, at low frequencies ($\nu\sim 1\,$GHz) the additional distortions
should reach $\sim 10\%$ of the intrinsic $y$-distortion, and a lower
frequency may even exceed it.
But in this context it is more important that due to the processes discussed
here new {\it narrow} spectral features appear, which have a unique
variability (e.g. see Fig.~\ref{fig:DT.25.diff_z}), that is even stronger
than in the case of the recombinational lines from standard recombination ($y=0$).
Such variability is very hard to mimic by any astrophysical foreground or
systematic problem with the instruments.  
As emphasized earlier for the cosmological recombination spectrum
\citep{RS2007, RS2008}, this may allow us to measure them {\it differentially},
also making use of the fact that practically the same signal is coming from
all directions on the sky.
For intrinsic $y$-distortions direct differential measurements are much
harder, since it has so broad spectrum.
Furthermore, as pointed out in the introduction, by measuring the narrow
spectral features under discussion one could in principle distinguish pre- and
post-recombinational energy release, an observation that cannot be achieved by
directly measuring the average $y$-distortion of the CMB.

\change{Above it was pointed out, that there is no principle difficulty in
  computing the spectral distortions due to pre-recombinational atomic
  transitions in \ion{H}{i} and \ion{He}{ii} for more general energy injection
  histories, if necessary. In particular very early injection, involving
  $\mu$-type distortions, may be interesting, since stimulated emission could
  strongly amplify the emission a low frequencies, and hence the total number
  of emitted photons per atom.
%
%
  However, to treat this problem one has to follow the detailed evolution of
  the CMB spectral distortion due to the injection process \citep[see
  e.g.][]{Hu1993}. Also the effect of electron scattering on the distortions
  due to the pre-recombinational atomic transitions, in particular because of
  the recoil effect, has to be treated more rigorously.
	Simple estimates show (see Appendix~\ref{app:ff_est}) that at low
	frequencies also the modifications due to the free-free absorption will
	become significant.
	Furthermore, one has to account for collisional processes, since they
  should become important at very high redshifts, even for shells with rather
  low $n$.

  An additional difficulty arises due to the fact that both at very low and at
  very high frequencies the back-reaction of the pre-recombinational
  distortion on the ambient radiation field may not be negligible (see
  Fig.~\ref{fig:comp_w_y}).
	This could also still affect the details of the results presented here,
	but the main conclusions should not change.
  We defer all these problems to another paper.  
}


\acknowledgement{ 
The authors wish to thank Jos\'e Alberto Rubi\~no-Mart\'{\i}n for useful
discussions.
We are also grateful for discussions on experimental possibilities with
J.~E.~Carlstrom, D.~J.~Fixsen, A.~Kogut, L.~Page, M.~Pospieszalski,
A.~Readhead, E.~J.~Wollack and especially J.~C.~Mather.
}

\begin{appendix}

\section{Computational details}
\label{app:comp}

\subsection{Formulation of the problem}
\label{app:problem}
Details on the formulation of the problem for a pure blackbody CMB radiation
field are given in \citet{Jose2006} and references therein. We shall use the
same notation here.
The main difference with respect to the standard recombination computation is
due to the possibility of a {\it non-blackbody} ambient radiation field, which
affects the net bound-bound and free-bound rates as explained in
Sect.~\ref{sec:atom_trans}.
Also the temperature of the electrons in general is no longer equal to the
effective temperature of the photons, as we discuss below (see
Sect.~\ref{app:rec_phot}).
Since we consider only small intrinsic spectral distortions all the
modification to the solution for the level populations are rather small, and
most of the differences will appear only as pre-recombinational emission due
to atomic transitions, \change{but with practically no effect on the
ionization history}.

One additional \change{modification} is related to the Lyman-continuum. As was
realized earlier \citep{Zeldovich68, Peebles68}, during the recombination
epochs photons cannot escape from the Lyman-continuum. However, at high
redshifts the number of neutral atoms is very small, so that the
Lyman-continuum becomes optically thin.
In order to include the escape of photons in the Lyman-continuum we follow the
analytic description of \citet{Chluba2007b}, in which an approximation for
escape probability in the Lyman-continuum was given by
\beal
\label{app:p_S_c_approx_b}
P^{\rm Ly-c}_{\rm esc}(z)\approx \frac{1}{1+\tau^{\rm esc}_{\rm c}},
\end{align}
with $\tau^{\rm esc}_{\rm c}=\frac{c \,\sigma_{{\rm 1s
c}}\,N_{1s}}{H}\frac{\kB\Te}{h\nu_{\rm c}}$. 
Here $\sigma_{{\rm 1s c}}$ is the threshold photoionization cross section of
the 1s-state, $N_{1s}$ is the number density of atoms in the ground state, and
$\nu_{\rm c}$ is the threshold frequency.
For the standard cosmology the \ion{H}{i} Lyman-continuum becomes optically
thin above $z\sim 3000-4000$, while for \ion{He}{ii} this occurs at $z\gtrsim
12000-16000$.
As our computations show, it is crucial to include this process, since at high
redshifts almost all loops begin or terminate in the Lyman-continuum (see
Sect.~\ref{sec:results}).

\subsection{High redshift solution}
\label{app:highz}
At high redshifts, well before the actual recombination epoch of the
considered atomic species, one can simplify the problem when realizing that
the ionization degree is actually not changing significantly. 
Although the inclusion of intrinsic CMB spectral distortion does lead to some
small changes in the populations with respect to the Saha values, the total
number of electrons that are captured by protons and helium nuclei is tiny as
compared to the total number of free electrons. Therefore one can neglect the
evolution equation for the electrons, until the actual recombination epoch is
entered. For \ion{H}{i} we use this simplification until $z\sim 3500$, while
for \ion{He}{ii} we follow the full system below $z\sim 20000$. Before we
simply use the {\sc Recfast}-solution for $\Ne$ \citep{SeagerRecfast1999,
Seager2000}. In several different cases we checked that these settings do not
affect the spectra.

Furthermore, one should mention that at high redshifts for $n>2$ we use the
variable $\Delta N_i=N_i-N^{\rm 2s}_i$ instead of $N_i$, since $\Delta
N_i/N_i$ becomes so small. Here $N^{\rm 2s}_i$ is the expected population of
level $i$ in Boltzmann-equilibrium relative to the 2s-level.
We then change back to the variable $N_i$ at sufficiently low redshifts.

\subsection{Recombination and photoionization rates}
\label{app:rec_phot}
The computation of the photoionization and recombination rates for many levels
is rather time-consuming. In an earlier version of our code \citep{Chluba2007}
we tabulated the recombination rates for all levels before the actual
computation and used {\it detailed balance} to infer the photoionization
rates.  This treatment is possible as long as the photon and electron
temperature do not depart significantly from each other, and when the
background spectrum is given by a blackbody. 
Here we now generalize this procedure in order to account for the small
difference in the electron and photon temperature, in particular at low
redshifts ($z\lesssim 800$), and to allow for non-blackbody ambient photon
distributions.

At high redshifts ($z\gtrsim 3000$) the electron temperature is always equal
to the Compton equilibrium temperature \citep{Zeldovich1969}:
\beal
\label{eq:Comp_eq}
T_{\rm e}^{\rm eq}=\frac{\kB\Tg}{h}\,\frac{\int x^4 n_\gamma(x) \id x}{4 \int x^3 n_\gamma(x) \id x}
\end{align}
within the given ambient radiation field. 
Due to the extremely high specific entropy of the Universe (there are
$\sim\pot{1.6}{9}$ photons per baryon) this temperature is reached on a much
shorter time-scale than the redistribution of photons via Compton scattering
requires.
For a $\mu$-type distortion $T_{\rm e}^{\rm eq}$ is always very close to the
effective photon temperature, whereas for a $y$-type distortion with $y\ll 1$
one has $\Te\approx \Tg[1+5.4\,y]$ \citep{Illarionov1975a}. This simplifies
matters, since there is no need to solve the electron temperature evolution
equation, and the photoionization and recombination rates can therefore be
pre-calculated.

At redshifts below $z\sim 3000$ we solve for the electron temperature
accounting for the non-blackbody ambient radiation field. In this case the
photon temperature inside the term due to the Compton interaction has to be
replaced by $T_{\rm e}^{\rm eq}$ as given by Eq.~\eqref{eq:Comp_eq}, such that
the temperature evolution equation reads
%
\beal
\label{eq:DT_Dz}
\pAb{\Te}{z}=\frac{\kappa_{\rm
C}\,T_\gamma^4}{H(z)[1+z]}\,\frac{X_{\rm e}}{1+f_{\rm He}+X_{\rm e}}
\,[\Te-T_{\rm e}^{\rm eq}]+\frac{2\Te}{1+z}
\Abst{,}
\end{align}
where $\kappa_{\rm C}=\pot{4.91}{-22}\,{\rm s^{-1}\,K^{-4}}$.

Since for small intrinsic CMB spectral distortions the correction to solution
for the temperature of the electrons is rather small, it is always possible to
use the standard {\sc Recfast} solution for $\Te$ as a reference. Tabulating
{\it both} the photoionization and recombination rates, and their first
derivatives with respect to the ratio of the electron to photon temperature
$\rho=\Te/\Tg$, it is possible to approximate the exact rates with high
accuracy using first order Taylor polynomials.
To save memory, we only consider all these rates in some range of redshifts
around the current point in the evolution and then update them from time to
time. At high redshifts we typically used 200 points per decade in logarithmic
spacing. At low redshift ($z\lesssim 5000$) we use 2 points per $\Delta z=1$.
Another improvement can be achieved by rescaling the reference solution for
$\Te$ with the true solution whenever the tabulated rates are updated.
With these settings we found excellent agreement with the full computation,
but at significantly lower computational cost.

\subsection{Inclusion of electron scattering}
\label{app:elec}
As has been mentioned by \citet{Dubrovich1997} and shown more detail
by \citet{Jose2007}, the broadening due to scattering of photons by free
electrons has to be included for the computation of the \ion{He}{ii}
recombination spectrum. Similarly, one has to account for this effect, when
computing the spectral distortions arising from higher redshifts.
Here we only consider redshifts $z\lesssim\pot{5}{4}$, and hence the electron
scattering Compton-$y$-parameter\footnote{Note that $\ye$ differs from $y$ as
defined in Eq.~\eqref{eq:y_par_Comp}, since it describes the redistribution of
some photon over frequency because of electron scattering rather than the
global energy exchange with the ambient blackbody radiation field.}
\beal
\label{eq:ye}
\ye(z)=\!\!\int_0^z\!\!\frac{k\Te}{\me c^2}\frac{c\,\Ne\sigT}{H(z')(1+z')}\id z'\approx\pot{4.8}{-11}\,[1+z]^2
\end{align}
is smaller than $\sim 0.12$, so that the line-broadening due to the
Doppler-effect is significant ($\Delta\nu/\nu|_{\rm Doppler}\sim 0.58$), but
still rather moderate in comparison with the width of the quasi-continuous
spectral features, arising from high redshifts (see
Sect.~\ref{sec:2shellresults} and \ref{sec:3shellresults}). However, already
at $z\lesssim\pot{2.5}{4}$ one has $\ye\lesssim 0.03$, such that
$\Delta\nu/\nu|_{\rm Doppler}\lesssim 0.29$.

Regarding the line-shifts due to the recoil-effect one finds that they are not
very important, since even for the \ion{H}{i} Lyman-continuum one has
$\Delta\nu/\nu|_{\rm recoil}\lesssim -0.14$ at $z\lesssim\pot{5}{4}$. Although
for the \ion{He}{ii} Lyman-continuum the shifts due to the recoil-effect is
four times larger, we shall not include it in our results. One therefore
expects that at frequencies $\nu \gtrsim 57\,$GHz the presented distortions
may still be modified due to this process, but we will consider this problem
in a future paper.

For the bound-bound spectrum we follow the procedure described in
\citet{Jose2007}, where the resulting spectral distortion at observing
frequency $\nu$ for one particular transition is given by
\citep[see also][]{Zeldovich1969}
\beal
\label{eq:DI_nu_Doppler}
\left.\Delta I_{ij}(\nu)\right|_{\rm Doppler}=
\int\frac{\nu^3}{\nu_0^3}\,
\frac{\Delta I_{ij}(\nu_0)}{\sqrt{4\pi\ye}}\times 
e^{\textstyle-\frac{(\ln[\nu/\nu_0]+3\ye)^2}{4\ye}}
\,\frac{\id\nu_0}{\nu_0}.
\end{align}
Here $\Delta I_{ij}(\nu_0)$ denotes the spectral distortion for the considered
transition evaluated at frequency $\nu_0$ and computed without the inclusion
of electrons scattering \citep[e.g. see][]{Jose2006}, but accounting for the
non-blackbody ambient radiation field. Note that $\ye(z)$ has to be calculated
starting at the emission redshift $\zem=\nu_{ij}/\nu_0-1$, where $\nu_{ij}$ is
the transition frequency.

For the spectral distortion resulting from the free-bound transitions one in
addition has to include the frequency-dependence of the photoionization cross
section.
We shall neglect the line broadening because of electrons scattering for the
moment.
Then, following \citet{Chluba2006b} and using the definitions of
Sect.~\ref{sec:fb_trans}, in the optically thin limit the spectral distortion
of the CMB at observing frequency $\nu$ due to direct recombinations to level
$i$ is given by
\beal
\label{eq:I_rec}
\Delta I_{i\rm c}(\nu)
&=
\frac{2h\nu^3}{c^2}\!
\int_{z_{\rm t}}^\infty 
n_\gamma(\nu_z, z)\,\frac{c N_i\,\sigma_{i}(\nu_z)}{H(z)(1+z)}
\nonumber
\\
&\qquad\times
\left[
\frac{\Ne\,N_{\rm p}}{N_{i}} \tilde{f}_i(\Te)\,e^{x_z+\mu(x_z)+[x_{i\rm c}-x_z]/\rho} -1
\right]\!
{\rm d} z,
\end{align}
with $\nu_z=\nu\,(1+z)$, $1+z_{\rm t}=\nu_{i{\rm c}}/\nu$ and
$x_z=h\nu_z/k\Tg\equiv h\nu/k T_0$. Furthermore, $n_\gamma(\nu_z, z)$ denotes
the intrinsic CMB occupation number at redshift $z$ including the spectral
distortion and evaluated at frequency $\nu_z$. For the Lyman-continuum one
would in addition multiply the integrand of Eq.~\eqref{eq:I_rec} by $P^{\rm
Ly-c}_{\rm esc}(z)$, to obtain the approximate solution for the resulting
distortion.

Now, to include the broadening because of scattering by electrons one has to
solve the 2-dimensional integral
\beal
\label{eq:I_rec_D}
\left.\Delta I_{i\rm c}(\nu)\right|_{\rm Doppler}
&=
\frac{2h\nu^3}{c^2}\!\!
\int_{0}^\infty \!\!\!\!\id z
\!\int
\frac{\Delta n(\tilde{\nu}_z, z)}{\sqrt{4\pi\ye}}
\, e^{\textstyle-\frac{(\ln[\nu/\tilde{\nu}]+3\ye)^2}{4\ye}}\frac{\id\tilde{\nu}}{\tilde{\nu}},
\end{align}
where $\tilde{\nu}_z=\tilde{\nu}\,(1+z)$ and
\beal
\label{eq:Delta_rec_D}
\Delta n(\nu, z)
\!&=\!
n_\gamma(\nu, z)\,\frac{c N_i\,\sigma_{i}(\nu)}{H(z)(1+z)}
\!\left[\!
\frac{\Ne\,N_{\rm p}}{N_{i}} \tilde{f}_i(\Te)\,e^{x+\mu(x)+[x_{i\rm c}-x]/\rho} -1\!
\right].
\end{align}
In the numerical evaluation of these integrals it is advisable to make use of
the knowledge about the integrand, since otherwise they may converge very
slowly.

\subsection{Estimate regarding the free-free process}
\label{app:ff_est}
The free-free optical depth, $\tau_{{\rm ff}}$, is given by
\beal
\label{eq:tau_ff}
\tau_{\rm ff}(x, z, z_{\rm f})=\int_{z}^{z_{\rm f}} \, K_{\rm ff}(x, z')\,
\frac{\Ne\,\sigT\,c\id z'}{H(z')(1+z')}
\Abst{,}
\end{align}
Here $\Ne$ is the free electron number density and $H(z)$ is the Hubble
factor, which in the radiation dominated era ($z\gtrsim 3300$) is given by
$H(z)\approx \pot{2.1}{-20}\,[1+z]^2\,{\rm s^{-1}}$. The free-free absorption
coefficient $K_{\rm ff}(x, z)$ is given by
\beal
\label{eq:K_ff}
K_{\rm ff}(x, z)=\frac{\alpha\,\lambda_{\rm
    e}^3}{2\pi\sqrt{6\pi}}\,\,
\frac{[1-e^{-x}]}{x^3\,\theta_{\gamma}^{7/2}}\,N_{\rm b}\,g_{\rm ff}^{\rm H^+}(x, \theta_{\gamma})
\Abst{,}
\end{align}
where $\lambda_{\rm e}=h/\me c= \pot{2.426}{-10}\,$cm is the Compton
wavelength of the electron, $\alpha\approx 1/137$ is the fine structure
constant, $g_{\rm ff}^{\rm H^+}$ is the free-free Gaunt factor for hydrogen.
We also introduced the dimensionless temperature of the photon field
$\theta_{\gamma}=\kB\Tg/\me c^2\approx\pot{4.6}{-10}\,[1+z]$.
For simplicity we again shall assume $\Te\equiv\Tg$. Furthermore we
  approximated the $\rm He^{++}$ free-free Gaunt factor by $g_{\rm ff}^{\rm
  He^{++}}\approx 4 g_{\rm ff, p} $ and assumed that $z\gtrsim 8000$, since in
  the considered frequency range most of the free-free absorption occurs well
  before $\ion{He}{iii}\rightarrow\ion{He}{ii}$ recombination (see below).

Using the condition $\tau_{\rm ff}\approx 1$ one can estimate the frequency
$x_{\rm ff}(z)$, below which one expects free-free absorption to become
important.
Since at $z\gtrsim 8000$ all the atoms are ionized, the number density of free
electrons is given by $\Ne=(1-Y_{\rm p}/2)\,N_{\rm b}\approx
\pot{2.2}{-7}\,(1+z)^3\,{\rm cm}^{-3}$. Here we used $N_{\rm
b,0}=\pot{2.5}{-7}\,{\rm cm}^{-3}$ as the present day baryon number density.
For $x_{\rm ff}\ll 1$ one then finds $\tau_{\rm ff}(x, z=0, z_{\rm
  em})\approx \pot{9.6}{-7}\,[g_{\rm ff}/5]\,\sqrt{1+z_{\rm em}}\,x^{-2}$,
  where $z_{\rm em}$ is the redshift of emission.
Here we are only interested in photons that can be observed at $x\gtrsim
0.02$, i.e. $\nu\gtrsim 1\,$GHz today.
At this frequency $\tau_{\rm ff}\gtrsim 1$ for $z_{\rm em}\gtrsim
\pot{1.7}{5}$. Below this redshift one can neglect the free-free process in
the computation of the bound-bound and free-bound spectra.
However, a more complete treatment will be presented in a future paper.

Assuming that $z_{\rm em}\sim 8000$ one finds that $\tau_{\rm ff}\lesssim
0.21$ for $x\gtrsim 0.02$. This justifies the approximations that were made
above, since the contributions to the free-free optical depth coming from
$z\lesssim 8000$ are not very large.


\section{Analytic solution}
\label{app:analytic}
\subsection{The 2-shell atom}
Including only 2 shells one can analytically derive the solution for the
Lyman-$\alpha$ line under quasi-stationary evolution of the populations.
For this we need to determine the net radiative rate, $\Delta R_{\rm
Ly_\alpha}=A_{21}(1+n_{21})\,N_{\rm 2p}[1-w N_{\rm 1s}/N_{\rm 2p}\times
n_{21}/(1+n_{21})]$, with $w=3$ and $n_{21}=n_\gamma(\nu_{21}, \Tg)$.
The ratio $\Lambda=n_{21}/(1+n_{21})$ is directly determined by the given
ambient radiation field including the spectral distortion. Since the
distortions are considered to be small we can use $A_{21}(1+n_{21})\,N_{\rm
2p}\approx A_{21}(1+n^{\rm eq}_{21})N^{\rm eq}_{\rm 2p}$ for the term in front
of the brackets. Here $n^{\rm eq}_{21}$ and $N^{\rm eq}_{\rm 2p}$ are
equilibrium values for the photon occupation number and the 2p-population,
respectively.
Therefore we only have to determine the ratio $\xi=w N_{\rm
1s}/N_{\rm 2p}$ in order to compute the Lyman-$\alpha$ line intensity
analytically.

We shall first consider the situation in the case of hydrogen at high
redshifts ($z\gtrsim 3000-4000$). There the escape probability in the
\ion{H}{i} Lyman-$\alpha$ line and the \ion{H}{i} Lyman-continuum are very
close to unity.  Therefore the 2s-1s-two-photon transition does not play any
important role in defining the number density of atoms in the ground
state. Furthermore, one can assume that the 2s-population is always in
Saha-equilibrium with the continuum, and hence
$N_{\rm 2s}\approx \Ne\Np \,\alpha_{\rm 2s}/\beta_{\rm 2s}$
where even $\Np\approx N_{\rm H}$, since the total fraction of neutral atoms
is tiny.

For the 1s- and 2p-states the rate equations read
\bsub
\label{eq:Rates_2shells}
\beal
&\Ne\Np \,\alpha_{\rm 1s}-\frac{\xi \beta_{\rm 1s}}{w}\,N_{\rm 2p}+A_{21}(1+n_{21})\,N_{\rm
2p}[1-\xi\Lambda]\approx 0
\\
&\Ne\Np \,\alpha_{\rm 2p}-\beta_{\rm 2p}\,N_{\rm 2p}-A_{21}(1+n_{21})\,N_{\rm 2p}[1-\xi\Lambda]\approx 0
\Abst{,}
\end{align}
\esub
where we substituted $N_{\rm 1s}=\xi N_{\rm 2p}/w$ and
$\Lambda=n_{21}/(1+n_{21})$. Solving this system with\footnote{Note that even
if one (more correctly) uses $\Np=N_{\rm H}-N_{\rm 1s}-N_{\rm 2s}-N_{\rm 2p}$
in Eq.~\eqref{eq:Rates_2shells} the solution for $\xi$ does not change.}
$\Np\approx N_{\rm H}$ for $\xi$ one finds
\beal
\label{eq:Sol_xi}
\xi=\frac{\alpha_{\rm 1s}\beta_{\rm 2p}+A_{21}(\alpha_{\rm 1s}+\alpha_{\rm
2p})[1+n_{21}]}{\alpha_{\rm 2p}\beta_{\rm 1s}/w+A_{21}(\alpha_{\rm 1s}+\alpha_{\rm
2p})\,n_{21}}
\Abst{.}
\end{align}
With appropriate replacements the same expression can be used to compute
the \ion{He}{ii} Lyman-$\alpha$ line.

\subsubsection{Including the Lyman-$\alpha$ and continuum escape}
\label{app:Tauc}
In order to include the escape probability in the Lyman-$\alpha$ line,
$P_{21}$, and the Lyman-continuum, $P_{\rm 1c}$, one simply should replace
$A_{21}\rightarrow P_{21}\,A_{21}$, $\alpha_{\rm 1s}\rightarrow P_{\rm
1c}\alpha_{\rm 1s}$ and $\beta_{\rm 1s}\rightarrow P_{\rm 1c}\beta_{\rm 1s}$,
where the escape probabilities can be computed using equilibrium values for
$N_{\rm 1s}$ and $N_{\rm 2p}$. As long as the 2s-1s-two-photon transition can
be neglected this yields a very accurate approximation for the Lyman-$\alpha$
line (cf. Sects.~\ref{sec:2shellresults}).

Around the region where the Lyman-continuum is becoming optically thick
($z\sim 3000$ for \ion{H}{i} and $z\sim 11000$ for \ion{He}{ii}), for simple
estimates one can use
\beal
\label{eq:tau_con}
\tau^{\rm esc}_{\rm Ly-c}&\approx
\begin{cases} 
\pot{7.2}{-24}\,e^{x_{\rm 1s}}\,[1+z]^{4} &\; \text{for \ion{H}{i}}
\\
\pot{3.1}{-24}\,e^{x_{\rm 1s}}\,[1+z]^{7/2} &\; \text{for \ion{He}{ii}}
\end{cases}
\\
\tau^{\rm esc}_{\rm Ly_\alpha}&\approx
\begin{cases} 
\pot{3.0}{-19}\,e^{x_{\rm 1s}}\,[1+z]^{3} &\; \text{for \ion{H}{i}}
\\
\pot{5.0}{-19}\,e^{x_{\rm 1s}}[1-e^{-3 x_{\rm 1s}/4}]\,[1+z]^{5/2} &\; \text{for \ion{He}{ii}}
\end{cases}
\Abst{,}
\end{align}
with $x_{\rm 1s}\approx \pot{5.79}{4}\,Z^2 [1+z]^{-1}$.

\subsubsection{More approximate behavior}
\label{app:moreapprox}
In order to understand the solution for the \ion{H}{i} Lyman-$\alpha$ line we
now turn to the corresponding intensity as a function of redshift
\citep[e.g.]{Jose2006}. This yields
\beal
\label{eq:Sol_DI}
\Delta I_\nu=\frac{h\,c}{4\pi}\,\frac{\Delta R_{\rm Ly_\alpha}(z)}{H(z)[1+z]^3}
=\frac{h\,c}{4\pi}\,\frac{A_{21}(1+n_{21})\,N_{\rm 2p}[1-\xi\Lambda]}{H(z)[1+z]^3}
\Abst{.}
\end{align}
Using the approximation~\eqref{eq:Sol_xi} for $\xi$ one can then find
\beal
\label{eq:1_xilam}
1-\xi\Lambda
\approx
\frac{\alpha_{\rm 2p}\beta_{\rm 1s}/w \,(1+n_{21})-\alpha_{\rm 1s}\beta_{\rm 2p}\,n_{21}}
{[\alpha_{\rm 2p}\beta_{\rm 1s}/w+A_{21}(\alpha_{\rm 1s}+\alpha_{\rm 2p})\,n_{21}](1+n_{21})}
\Abst{.}
\end{align}
With this one then has
\beal
\label{eq:Sol_DI_b}
%
\Delta I_\nu
&\approx\frac{h\,c}{4\pi}\,\frac{A_{21}\,N_{\rm 2p}}{H(z)[1+z]^3}
\,
\frac{\alpha_{\rm 2p}\beta_{\rm 1s} \,(1+n_{21})}
{\alpha_{\rm 2p}\beta_{\rm 1s}+w\,A_{21}(\alpha_{\rm 1s}+\alpha_{\rm 2p})\,n_{21}}
\nonumber\\
&\quad\quad\quad\quad\times
\left[1-w\,\frac{\alpha_{\rm 1s}\beta_{\rm 2p}}{\alpha_{\rm 2p}\beta_{\rm 1s}}\,e^{-(x_{21}+\mu_{21})}\right]
\Abst{.}
\end{align}
Here we used $x_{21}=h\nu_{21}/k\Tg$ and
$(1+n_{21})/n_{21}=e^{x_{21}+\mu_{21}}$, with the frequency dependent chemical
potential $\mu_{21}=\mu(x_{21})$.

\subsubsection*{First factor}
We can now simplify the expression \eqref{eq:Sol_DI_b} when realizing that
except for the term inside brackets, in the case of small intrinsic CMB
spectral distortions, one can just use equilibrium values.
At high redshifts one has $H(z)\propto (1+z)^2$. Furthermore $N^{\rm eq}_{\rm
2p}\approx 3\, N^{\rm eq}_{\rm 1s}\,e^{-x_{21}}\approx 3\,N_{\rm e}\,N_{\rm
H}\,\alpha^{\rm eq}_{\rm 1s}\,e^{-x_{21}}/\beta^{\rm eq}_{\rm 1s}$ and
$\beta^{\rm eq}_{i}=\alpha^{\rm eq}_{i}\,e^{-\xic}/\tilde{f}_i(\Te)$. Also
with rather high accuracy one finds $\alpha^{\rm eq}_{\rm
2p}\approx\alpha^{\rm eq}_{\rm 1s}/3$ and $\frac{\alpha_{\rm 2p}\beta_{\rm
1s}}{w}\ll A_{21}(\alpha_{\rm 1s}+\alpha_{\rm 2p})\,n_{21}$, so that
\beal
\label{eq:factor_a}
F(z)&=\frac{h\,c}{4\pi}\,\frac{A_{21}\,N_{\rm 2p}}{H(z)[1+z]^3}
\,
\frac{\alpha_{\rm 2p}\beta_{\rm 1s} \,(1+n_{21})}
{\alpha_{\rm 2p}\beta_{\rm 1s}+w\,A_{21}(\alpha_{\rm 1s}+\alpha_{\rm 2p})\,n_{21}}
\nonumber\\
&\quad\quad
\approx
\frac{h\,c}{4\pi}\,\frac{3\,N_{\rm e}\,N_{\rm H}}{H(z)[1+z]^3}
\,
\frac{\alpha^{\rm eq}_{\rm 2p}}{4+\tau^{\rm esc}_{\rm Ly-c}} \propto 
\frac{(1+z)^{1/2}}{4+\tau^{\rm esc}_{\rm Ly-c}}
\Abst{.}
\end{align}
Here we have also included the escape probabilities in the Lyman-$\alpha$ line
and continuum as explained in Appendix~\ref{app:Tauc}. Note that the
Lyman-$\alpha$ escape probability drops out of the expression, so that only
the Lyman-continuum escape probability is strongly affecting the
pre-recombinational line shape.
We have also used $\alpha^{\rm eq}_{i}=\frac{8\pi}{c^2}
\tilde{f}_i(\Te)\,e^{\xic}\,I^{\rm eq}_i$, with the integral
\bsub
\label{eq:Iint}
\beal
I^{\rm eq}_{i}&=\int_{\nuic}^\infty
\frac{\nu^2\,\sigma_i(\nu)}{e^{x}-1}\id\nu
\approx
\sigma_i(\nuic)\,\nu^3_{i\rm c} 
\,M_{-1}(\xic)
\Abst{,}
\end{align}
\esub
where we have assumed $\sigma_i(\nu)\approx \sigma_i(\nuic) \frac{\nu^3_{i\rm
c}}{\nu^3}$. The integral $M_i(x)$ is defined and discussed in
Appendix~\ref{app:someInts}.
For the 2p-state one has $\sigma_{\rm 2p}(\nu_{\rm 2pc})\,\nu^3_{\rm
2pc}\approx \pot{7.54}{27}\,Z^4\,{\rm cm^{-2}\,s^{-3}}$.

We checked the scaling of $F$ numerically and found 
\beal
\label{eq:factor_approx}
F(z)&\approx
\pot{5.6}{-26}\,\frac{(1+z)^{1/2}}{1+\tau^{\rm esc}_{\rm Ly-c}/4}\,{\rm J\,m^{-2}\,s^{-1}\,Hz^{-1}\,sr^{-1}}
\end{align}
within $\lesssim 20\%$ accuracy in the important redshift range.

\subsubsection*{Second factor}
Using the definitions of $\alpha_{i}$ and $\beta_{i}$ as given in
Sect.~\ref{sec:bb_trans} and \ref{sec:fb_trans} one directly finds
$w\,\frac{\alpha_{\rm 1s}\beta_{\rm 2p}}{\alpha_{\rm 2p}\beta_{\rm
1s}}\,e^{-(x_{21}+\mu_{21})}\equiv e^{-\mu_{21}}\,G(z)$
with
\beal
\label{eq:G}
G(z)&=\frac{
\sigav{n\,e^{\mu(x)+(\xonesc-x)\Delta\rho/\rho} }{\rm 1s}
\sigav{n}{\rm 2p}}
{\sigav{n\,e^{\mu(x)+(\xtwopc-x)\Delta\rho/\rho}}{\rm 2p} \sigav{n}{\rm 1s}}
\Abst{.}
\end{align}
Here $\Delta\rho=1-\rho$ and we introduced the notation
\beal
\label{eq:sigav}
\sigav{f(\nu)}{i}=\int_{\xic}^\infty \nu^2 \sigma_i(\nu) f(\nu)\id\nu
\end{align}
for the average of some function $f(\nu)$ over the photoionization
cross-section of level $i$.

In full thermodynamic equilibrium one has $G^{\rm eq}(z)\equiv 1$, a property
that can be verified using Eq.~\eqref{eq:G} with $\mu=0$ and $\rho=1$, since
then $\sigav{n\,e^{\mu(x)+(\xic-x)\Delta\rho/\rho} }{\rm i}\equiv
\sigav{n_{\rm pl}}{\rm i}$.
Therefore we can write $G=1+\Delta G$. 
Using $\sigav{f}{i}=\sigav{f^{\rm eq}}{i}+\sigav{\Delta f}{i}$, for small
intrinsic CMB distortions (i.e. $\sigav{\Delta f}{i}/\sigav{f}{i}\ll 1$) one finds
\beal 
\label{eq:DG}
\Delta G
\approx
 \frac{\sigav{n_{\rm pl}\,[\mu-\mu^{\rho}_{\rm 1s}]}{\rm 1s}}
{\sigav{n_{\rm pl}}{\rm 1s}}
-\frac{\sigav{n_{\rm pl}\,[\mu-\mu^\rho_{\rm 2p}]}{\rm 2p}}
{\sigav{n_{\rm pl}}{\rm 2p}}
\Abst{,}
\end{align}
where $\mu^\rho_{i}=(x-\xic)\Delta\rho/\rho$.
Putting things together we then have
\bsub
\label{eq:imbal}
\beal
1-&w\,\frac{\alpha_{\rm 1s}\beta_{\rm 2p}}{\alpha_{\rm 2p}\beta_{\rm
1s}}\,e^{-(x_{21}+\mu_{21})}
\nonumber\\
\label{eq:imbal_a}
&\approx \mu_{21}
+\frac{\sigav{n_{\rm pl}\,\mu}{\rm 2p}}{\sigav{n_{\rm pl}}{\rm 2p}}
-\frac{\sigav{n_{\rm pl}\,\mu}{\rm 1s}}{\sigav{n_{\rm pl}}{\rm 1s}}
\nonumber\\
&\qquad 
-\frac{\sigav{n_{\rm pl}\,\mu^\rho_{\rm 2p}}{\rm 2p}}{\sigav{n_{\rm pl}}{\rm 2p}}
+\frac{\sigav{n_{\rm pl}\,\mu^\rho_{\rm 1s}}{\rm 1s}}{\sigav{n_{\rm pl}}{\rm 1s}}
\\
\label{eq:imbal_b}
&\approx 
\mu_{21}+\mu_{\rm 2pc}-\mu_{\rm 1sc}
\Abst{,}
\end{align}
\esub
To lowest order, Eq. \eqref{eq:imbal_b} shows that the main reason for the
emission in the Lyman-$\alpha$ line is the deviation of the effective chemical
potential from zero at the Lyman-$\alpha$ resonance, and the Lyman- and
Balmer-continuum frequency. However, the averages over the photoionization
cross-section still lead to some notable corrections, so that also the small
difference in the electron and photon temperature plays a role.

If we again use the Kramers-approximation for the photoionization
cross-section, $\sigma_i(\nu)\approx \sigma_i(\nuic) \frac{\nu^3_{i\rm
c}}{\nu^3}$, looking at Eq.~\eqref{eq:mu_n} for $\mu$ in the case of a small
$y$-type distortion, one can write
\bsub
\label{eq:approx_sigav}
\beal
\sigav{n_{\rm pl}}{i}&\approx \kappa_i\,M_{-1}(\xic)
\\
\sigav{n_{\rm pl}\,\mu}{i}&\approx \kappa_i\,y\,[4\,M_0(\xic)-S(\xic)]
\\
\sigav{n_{\rm pl}\,\mu^\rho_{i}}{i}&\approx \kappa_i\,\frac{\Delta\rho}{\rho}[M_0(\xic)-\xic\,M_{-1}(\xic)]
\Abst{,}
\end{align}
\esub
where $\kappa_i=\rm const$ and the integrals $S$ and $M_0$ are defined in
Appendix~\ref{app:someInts}.
Keeping only the leading order terms, we have
\bsub
\label{eq:imbal_approx}
\beal
1-&w\,\frac{\alpha_{\rm 1s}\beta_{\rm 2p}}{\alpha_{\rm 2p}\beta_{\rm
1s}}\,e^{-(x_{21}+\mu_{21})}
\nonumber\\
&\approx 
-y\,x_{\rm 1sc}\left[6.3-0.9375\,x_{\rm 1sc}+4.7\,e^{-x_{\rm 1sc}}-1.175\,e^{-x_{\rm 1sc}/4} \right]
\nonumber\\
&\qquad\qquad
+y\,x_{21}\left[ 9.4-x_{21}\,\frac{e^{x_{21}}+1}{e^{x_{21}}-1} \right]
\Abst{.}
\end{align}
\esub
It is important to mention that this is still a rather rough approximation,
since already applying the Kramers-formula for the photoionization
cross-section introduces some significant simplification.
However, this approximation may be useful for simple estimates.

\section{Some integrals}
\label{app:someInts}
\subsection{Integrals $M_{i}$}
In the evaluation of the recombination and photoionization rates, integrals of
the form $M_i=\int_\xic^\infty x^i\id x/[e^x-1]$ appear. Below we now discuss
those which are of importance for us here.

\subsubsection{Integral $M_{-1}$}
For $i=-1$ one can write
\bsub
\label{eq:Iint_b}
\beal 
\int_{\xic}^\infty \frac{\id x}{x[e^{x}-1]} &= \sum_{k=1}^{\infty} {\rm Ei}(k\xic)
\label{eq:Iint_b_1}
\stackrel{\stackrel{h\nuic\gg k\Tg}{\downarrow}}{\approx}
\frac{e^{-\xic}}{\xic} \\
\label{eq:Iint_b_2}
&\!\!\!
\!\!\!\!\!\!\!\!\!\!\!\!\!\!\!\!\!\!\!\!\!\!\!\!
\stackrel{\stackrel{h\nuic\leq k\Tg}{\downarrow}}{\approx}
\frac{1}{\xic}\left[1-\frac{11-6\gamma}{12}\,\xic-\frac{x^2_{i\rm
c}}{12}+\frac{\xic}{2}\ln(\xic)\right] \Abst{,}
\end{align}
\esub
where $\gamma\approx 0.5772$ is the Euler constant and we made used of the
exponential integral ${\rm Ei}(x)=\int_{x}^\infty e^{-t}\id t/t$.
In the limit $h\nuic\leq k\Tg$ the given approximation is accurate to better
than 1\%. For $h\nuic\geq k\Tg$ the first five terms in the full sum also
yield similar accuracy.

Since $\xic\approx \pot{5.79}{4}\,Z^2 n^{-2} [1+z]^{-1}$, it is clear that at
$z\lesssim \pot{1.45}{4}\,Z^2$ both Lyman- and Balmer-continuum are still in
the exponential tail of the CMB blackbody. In the redshift range
$\pot{1.45}{4}\,Z^2\lesssim z \lesssim \pot{5.79}{4}\,Z^2$ the Lyman-continuum
is still in the exponential tail of the CMB, while the Balmer-continuum is
already in the Rayleigh-Jeans part of the spectrum. Only at $z\gtrsim
\pot{5.79}{4}\,Z^2$ one can use the low frequency expansion of
Eq.~\eqref{eq:Iint} for both cases.
However, to within $\lesssim 30\%$ one may also apply Eq. \eqref{eq:Iint_b_1}
in the whole range.

\subsubsection{Integral $M_{0}$}
For $i=0$ one can write
\beal
\label{eq:M0}
M_0&=\int_\xic^\infty \id x/[e^x-1]=\sum_{k=1}^\infty \int_\xic^\infty
e^{-kx}\id x=\sum_{k=1}^\infty e^{-k\xic}/k 
\nonumber\\
&= \xic-\ln(e^\xic-1) 
\Abst{,}
\end{align}
which for $\xic\lesssim 1$ can be approximated as $M_0 \approx
\frac{\xic}{2}-\frac{x_{i\rm c}^2}{24}-\ln(\xic)$, while for $\xic\gg 1$ one
has $M_0 \approx e^{-\xic}\,[1+e^{-\xic}/2]$.


\subsection{Integral $S$}
In the evaluation of the recombination and photoionization rates one also
encounters $S(x)=\int_\xic^\infty \id x\, x \,\frac{e^x+1}{[e^x-1]^2}$. 
The first part of this integral, $\propto x e^x/[e^x-1]^2$, can be directly
taken yielding $\int_\xic^\infty \id x\, x e^x/[e^x-1]^2=\xic
e^\xic/[e^\xic-1]-\ln(e^\xic-1)$.
Introducing the polylogarithm ${\rm Li}_n(x)=\sum_{k=1}^\infty x^k/k^n$ and
realizing $x /[e^x-1]^2=\sum_{k=1}^\infty k\,x\,e^{-(k+1)\,x}$ one can find
\bsub
\label{eq:S}
\beal
S(\xic)&=\xic \,\frac{e^\xic+1}{e^\xic-1}+\xic(1-\xic)
\nonumber\\
&\qquad\qquad-(2-\xic)\ln(e^\xic-1)-{\rm Li}_2(e^{-\xic})
\nonumber\\
\label{eq:eq:S_1}
&\!\!\!
\stackrel{\stackrel{h\nuic\gtrsim k\Tg}{\downarrow}}{\approx}
\sum_{k=1}^{m\approx 5} \frac{2k-1}{k^2}\left[1+k\,\xic\right]\,e^{-k\xic}
\\
\label{eq:eq:S_2}
&\!\!\!
\stackrel{\stackrel{h\nuic\lesssim k\Tg}{\downarrow}}{\approx}
2-\frac{\pi^2}{6}+\xic-\frac{x^2_{i\rm c}}{6}-2\ln(\xic)
\Abst{,}
\end{align}
\esub
The given approximations are accurate to $\lesssim 1\%$.

\end{appendix}

\bibliographystyle{aa} 
\bibliography{Lit}

\begin{thebibliography}{58}
\expandafter\ifx\csname natexlab\endcsname\relax\def\natexlab#1{#1}\fi

\bibitem[{{Bennett} {et~al.}(2003){Bennett}, {Halpern}, {Hinshaw}, {Jarosik},
  {Kogut}, {Limon}, {Meyer}, {Page}, {Spergel}, {Tucker}, {Wollack}, {Wright},
  {Barnes}, {Greason}, {Hill}, {Komatsu}, {Nolta}, {Odegard}, {Peiris}, \&
  {Verde}}]{WMAP_params}
{Bennett}, C.~L., {Halpern}, M., {Hinshaw}, G., {et~al.} 2003, \apjs, 148, 1

\bibitem[{{Burigana} {et~al.}(1991{\natexlab{a}}){Burigana}, {Danese}, \& {de
  Zotti}}]{Burigana1991b}
{Burigana}, C., {Danese}, L., \& {de Zotti}, G. 1991{\natexlab{a}}, \apj, 379,
  1

\bibitem[{{Burigana} {et~al.}(1991{\natexlab{b}}){Burigana}, {Danese}, \& {de
  Zotti}}]{Burigana1991}
{Burigana}, C., {Danese}, L., \& {de Zotti}, G. 1991{\natexlab{b}}, \aap, 246,
  49

\bibitem[{{Burigana} {et~al.}(1995){Burigana}, {de Zotti}, \&
  {Danese}}]{Burigana1995}
{Burigana}, C., {de Zotti}, G., \& {Danese}, L. 1995, \aap, 303, 323

\bibitem[{{Burigana} \& {Salvaterra}(2003)}]{Burigana2003}
{Burigana}, C. \& {Salvaterra}, R. 2003, \mnras, 342, 543

\bibitem[{{Cen} \& {Ostriker}(1999)}]{Cen1999}
{Cen}, R. \& {Ostriker}, J.~P. 1999, \apj, 514, 1

\bibitem[{{Chan} {et~al.}(1975){Chan}, {Grant}, \& {Jones}}]{Chan1975}
{Chan}, K.~L., {Grant}, C., \& {Jones}, B.~J.~T. 1975, \apj, 195, 1

\bibitem[{{Chluba} {et~al.}(2007{\natexlab{a}}){Chluba},
  {Rubi{\~n}o-Mart{\'{\i}}n}, \& {Sunyaev}}]{Chluba2007}
{Chluba}, J., {Rubi{\~n}o-Mart{\'{\i}}n}, J.~A., \& {Sunyaev}, R.~A.
  2007{\natexlab{a}}, \mnras, 374, 1310

\bibitem[{{Chluba} {et~al.}(2007{\natexlab{b}}){Chluba}, {Sazonov}, \&
  {Sunyaev}}]{Chluba2007DC}
{Chluba}, J., {Sazonov}, S.~Y., \& {Sunyaev}, R.~A. 2007{\natexlab{b}}, \aap,
  468, 785

\bibitem[{{Chluba} \& {Sunyaev}(2004)}]{Chluba2004}
{Chluba}, J. \& {Sunyaev}, R.~A. 2004, \aap, 424, 389

\bibitem[{{Chluba} \& {Sunyaev}(2006)}]{Chluba2006b}
{Chluba}, J. \& {Sunyaev}, R.~A. 2006, \aap, 458, L29

\bibitem[{{Chluba} \& {Sunyaev}(2007)}]{Chluba2007b}
{Chluba}, J. \& {Sunyaev}, R.~A. 2007, \aap, 475, 109

\bibitem[{{Chluba} \& {Sunyaev}(2008)}]{Chluba2008}
{Chluba}, J. \& {Sunyaev}, R.~A. 2008, \aap, 478, L27

\bibitem[{{da Silva} {et~al.}(2000){da Silva}, {Barbosa}, {Liddle}, \&
  {Thomas}}]{Silva2000}
{da Silva}, A.~C., {Barbosa}, D., {Liddle}, A.~R., \& {Thomas}, P.~A. 2000,
  \mnras, 317, 37

\bibitem[{{Daly}(1991)}]{Daly1991}
{Daly}, R.~A. 1991, \apj, 371, 14

\bibitem[{{Danese} \& {de Zotti}(1982)}]{Danese1982}
{Danese}, L. \& {de Zotti}, G. 1982, \aap, 107, 39

\bibitem[{{Dubrovich}(1975)}]{Dubrovich1975}
{Dubrovich}, V.~K. 1975, Soviet Astronomy Letters, 1, 196

\bibitem[{{Dubrovich} \& {Stolyarov}(1995)}]{DubroVlad95}
{Dubrovich}, V.~K. \& {Stolyarov}, V.~A. 1995, \aap, 302, 635

\bibitem[{{Dubrovich} \& {Stolyarov}(1997)}]{Dubrovich1997}
{Dubrovich}, V.~K. \& {Stolyarov}, V.~A. 1997, Astronomy Letters, 23, 565

\bibitem[{{Fixsen} {et~al.}(1996){Fixsen}, {Cheng}, {Gales}, {Mather},
  {Shafer}, \& {Wright}}]{Fixsen1996}
{Fixsen}, D.~J., {Cheng}, E.~S., {Gales}, J.~M., {et~al.} 1996, \apj, 473, 576

\bibitem[{{Fixsen} {et~al.}(1998){Fixsen}, {Dwek}, {Mather}, {Bennett}, \&
  {Shafer}}]{Fixsen1998}
{Fixsen}, D.~J., {Dwek}, E., {Mather}, J.~C., {Bennett}, C.~L., \& {Shafer},
  R.~A. 1998, \apj, 508, 123

\bibitem[{{Fixsen} \& {Mather}(2002)}]{Fixsen2002}
{Fixsen}, D.~J. \& {Mather}, J.~C. 2002, \apj, 581, 817

\bibitem[{{Hu} {et~al.}(1994){Hu}, {Scott}, \& {Silk}}]{Hu1994}
{Hu}, W., {Scott}, D., \& {Silk}, J. 1994, \apjl, 430, L5

\bibitem[{{Hu} \& {Silk}(1993{\natexlab{a}})}]{Hu1993}
{Hu}, W. \& {Silk}, J. 1993{\natexlab{a}}, \prd, 48, 485

\bibitem[{{Hu} \& {Silk}(1993{\natexlab{b}})}]{Hu1993a}
{Hu}, W. \& {Silk}, J. 1993{\natexlab{b}}, Physical Review Letters, 70, 2661

\bibitem[{{Illarionov} \& {Syunyaev}(1975{\natexlab{a}})}]{Illarionov1975a}
{Illarionov}, A.~F. \& {Syunyaev}, R.~A. 1975{\natexlab{a}}, Soviet Astronomy,
  18, 413

\bibitem[{{Illarionov} \& {Syunyaev}(1975{\natexlab{b}})}]{Illarionov1975b}
{Illarionov}, A.~F. \& {Syunyaev}, R.~A. 1975{\natexlab{b}}, Soviet Astronomy,
  18, 691

\bibitem[{{Kaplan} \& {Pikelner}(1970)}]{Kaplan1970}
{Kaplan}, S.~A. \& {Pikelner}, S.~B. 1970, {The interstellar medium}
  (Cambridge: Harvard University Press, 1970)

\bibitem[{{Kogut} {et~al.}(2006){Kogut}, {Fixsen}, {Fixsen}, {Levin}, {Limon},
  {Lowe}, {Mirel}, {Seiffert}, {Singal}, {Lubin}, \& {Wollack}}]{Kogut2006}
{Kogut}, A., {Fixsen}, D., {Fixsen}, S., {et~al.} 2006, New Astronomy Review,
  50, 925

\bibitem[{{Kogut} {et~al.}(2004){Kogut}, {Fixsen}, {Levin}, {Limon}, {Lubin},
  {Mirel}, {Seiffert}, \& {Wollack}}]{Kogut2004}
{Kogut}, A., {Fixsen}, D.~J., {Levin}, S., {et~al.} 2004, \apjs, 154, 493

\bibitem[{{Lagache} {et~al.}(2005){Lagache}, {Puget}, \& {Dole}}]{Lagache2005}
{Lagache}, G., {Puget}, J.-L., \& {Dole}, H. 2005, \araa, 43, 727

\bibitem[{{Lightman}(1981)}]{Lightman1981}
{Lightman}, A.~P. 1981, \apj, 244, 392

\bibitem[{{Lyubarsky} \& {Sunyaev}(1983)}]{Liubarskii83}
{Lyubarsky}, Y.~E. \& {Sunyaev}, R.~A. 1983, \aap, 123, 171

\bibitem[{{Markevitch} {et~al.}(1991){Markevitch}, {Blumenthal}, {Forman},
  {Jones}, \& {Sunyaev}}]{Markevitch1991}
{Markevitch}, M., {Blumenthal}, G.~R., {Forman}, W., {Jones}, C., \& {Sunyaev},
  R.~A. 1991, \apjl, 378, L33

\bibitem[{{Mather} {et~al.}(1999){Mather}, {Fixsen}, {Shafer}, {Mosier}, \&
  {Wilkinson}}]{Mather1999}
{Mather}, J.~C., {Fixsen}, D.~J., {Shafer}, R.~A., {Mosier}, C., \&
  {Wilkinson}, D.~T. 1999, \apj, 512, 511

\bibitem[{{McDonald} {et~al.}(2001){McDonald}, {Scherrer}, \&
  {Walker}}]{McDonald2001}
{McDonald}, P., {Scherrer}, R.~J., \& {Walker}, T.~P. 2001, \prd, 63, 023001

\bibitem[{{Miniati} {et~al.}(2000){Miniati}, {Ryu}, {Kang}, {Jones}, {Cen}, \&
  {Ostriker}}]{Miniati2000}
{Miniati}, F., {Ryu}, D., {Kang}, H., {et~al.} 2000, \apj, 542, 608

\bibitem[{{Oh} {et~al.}(2003){Oh}, {Cooray}, \& {Kamionkowski}}]{Oh2003}
{Oh}, S.~P., {Cooray}, A., \& {Kamionkowski}, M. 2003, \mnras, 342, L20

\bibitem[{{Peebles}(1968)}]{Peebles68}
{Peebles}, P.~J.~E. 1968, \apj, 153, 1

\bibitem[{{Roncarelli} {et~al.}(2007){Roncarelli}, {Moscardini}, {Borgani}, \&
  {Dolag}}]{Roncarelli2007}
{Roncarelli}, M., {Moscardini}, L., {Borgani}, S., \& {Dolag}, K. 2007, \mnras,
  378, 1259

\bibitem[{{Rubi{\~n}o-Mart{\'{\i}}n} {et~al.}(2006){Rubi{\~n}o-Mart{\'{\i}}n},
  {Chluba}, \& {Sunyaev}}]{Jose2006}
{Rubi{\~n}o-Mart{\'{\i}}n}, J.~A., {Chluba}, J., \& {Sunyaev}, R.~A. 2006,
  \mnras, 371, 1939

\bibitem[{{Rubino-Martin} {et~al.}(2007){Rubino-Martin}, {Chluba}, \&
  {Sunyaev}}]{Jose2007}
{Rubino-Martin}, J.~A., {Chluba}, J., \& {Sunyaev}, R.~A. 2007, ArXiv e-prints,
  711

\bibitem[{{Salvaterra} \& {Burigana}(2002)}]{Salvaterra2002}
{Salvaterra}, R. \& {Burigana}, C. 2002, \mnras, 336, 592

\bibitem[{{Seager} {et~al.}(1999){Seager}, {Sasselov}, \&
  {Scott}}]{SeagerRecfast1999}
{Seager}, S., {Sasselov}, D.~D., \& {Scott}, D. 1999, \apjl, 523, L1

\bibitem[{{Seager} {et~al.}(2000){Seager}, {Sasselov}, \& {Scott}}]{Seager2000}
{Seager}, S., {Sasselov}, D.~D., \& {Scott}, D. 2000, \apjs, 128, 407

\bibitem[{{Spergel} {et~al.}(2003){Spergel}, {Verde}, {Peiris}, {Komatsu},
  {Nolta}, {Bennett}, {Halpern}, {Hinshaw}, {Jarosik}, {Kogut}, {Limon},
  {Meyer}, {Page}, {Tucker}, {Weiland}, {Wollack}, \& {Wright}}]{Spergel2003}
{Spergel}, D.~N., {Verde}, L., {Peiris}, H.~V., {et~al.} 2003, \apjs, 148, 175

\bibitem[{{Sunyaev} \& {Chluba}(2007)}]{RS2007}
{Sunyaev}, R.~A. \& {Chluba}, J. 2007, ArXiv e-prints, 710

\bibitem[{{Sunyaev} \& {Chluba}(2008)}]{RS2008}
{Sunyaev}, R.~A. \& {Chluba}, J. 2008, ArXiv e-prints, 802

\bibitem[{{Sunyaev} \& {Zeldovich}(1980)}]{Sunyaev1980ARAA}
{Sunyaev}, R.~A. \& {Zeldovich}, I.~B. 1980, \araa, 18, 537

\bibitem[{{Sunyaev} \& {Zeldovich}(1970{\natexlab{a}})}]{Sunyaev1970Antimatter}
{Sunyaev}, R.~A. \& {Zeldovich}, Y.~B. 1970{\natexlab{a}}, \apss, 9, 368

\bibitem[{{Sunyaev} \& {Zeldovich}(1970{\natexlab{b}})}]{Sunyaev1970b}
{Sunyaev}, R.~A. \& {Zeldovich}, Y.~B. 1970{\natexlab{b}}, \apss, 7, 20

\bibitem[{{Sunyaev} \& {Zeldovich}(1970{\natexlab{c}})}]{Sunyaev1970COA}
{Sunyaev}, R.~A. \& {Zeldovich}, Y.~B. 1970{\natexlab{c}}, Comments on
  Astrophysics and Space Physics, 2, 66

\bibitem[{{Sunyaev} \& {Zeldovich}(1972{\natexlab{a}})}]{Sunyaev1972b}
{Sunyaev}, R.~A. \& {Zeldovich}, Y.~B. 1972{\natexlab{a}}, \aap, 20, 189

\bibitem[{{Sunyaev} \& {Zeldovich}(1972{\natexlab{b}})}]{Sunyaev1972}
{Sunyaev}, R.~A. \& {Zeldovich}, Y.~B. 1972{\natexlab{b}}, Comments on
  Astrophysics and Space Physics, 4, 173

\bibitem[{{Thorne}(1981)}]{Thorne1981}
{Thorne}, K.~S. 1981, \mnras, 194, 439

\bibitem[{{Zeldovich} {et~al.}(1972){Zeldovich}, {Illarionov}, \&
  {Syunyaev}}]{Zeldovich1972}
{Zeldovich}, Y.~B., {Illarionov}, A.~F., \& {Syunyaev}, R.~A. 1972, Soviet
  Journal of Experimental and Theoretical Physics, 35, 643

\bibitem[{{Zeldovich} {et~al.}(1968){Zeldovich}, {Kurt}, \&
  {Syunyaev}}]{Zeldovich68}
{Zeldovich}, Y.~B., {Kurt}, V.~G., \& {Syunyaev}, R.~A. 1968, Zhurnal
  Eksperimental noi i Teoreticheskoi Fiziki, 55, 278

\bibitem[{{Zeldovich} \& {Sunyaev}(1969)}]{Zeldovich1969}
{Zeldovich}, Y.~B. \& {Sunyaev}, R.~A. 1969, \apss, 4, 301

\end{thebibliography}

\end{document}